\documentclass[aps,prx,twocolumn,showpacs,amsmath,amssymb,floatfix,longbibliography,footinbib,10pt]{revtex4-2}
\usepackage{graphicx}
\usepackage{mathtools,bm,array,enumitem,overpic,booktabs,multirow}
\usepackage{natbib}
\usepackage[dvipsnames]{xcolor}
\usepackage{braket}
\usepackage{float}
\usepackage{textcomp}
\usepackage{comment}
\usepackage[bookmarks=true,colorlinks,linkcolor=RoyalBlue,urlcolor=RoyalBlue,citecolor=RoyalBlue]{hyperref}
\usepackage{cleveref}
\usepackage{hhline}
\usepackage[normalem]{ulem}
\usepackage{orcidlink}
\begin{document}
\title{Coherent Dynamics in Chaotic Spin Chains via Interference-Protected Subspaces}
\author{Aron Kerschbaumer${}^{\orcidlink{0009-0002-2370-8661}}$}
\email{aron.kerschbaumer@ist.ac.at}
\affiliation{Institute of Science and Technology Austria (ISTA), Am Campus 1, 3400 Klosterneuburg, Austria}
\author{Jean-Yves Desaules${}^{\orcidlink{0000-0002-3749-6375}}$}
\email{jean-yves.desaules@ist.ac.at}
\affiliation{Institute of Science and Technology Austria (ISTA), Am Campus 1, 3400 Klosterneuburg, Austria}
\author{Maksym Serbyn${}^{\orcidlink{0000-0002-2399-5827}}$}
\email{maksym.serbyn@ist.ac.at}
\affiliation{Institute of Science and Technology Austria (ISTA), Am Campus 1, 3400 Klosterneuburg, Austria}
\date{\today}

\begin{abstract}
Generic quantum many-body systems are expected to thermalize, scrambling initial coherence while local observables relax to equilibrium values. Weak ergodicity breaking, often associated with scarring of homogeneous states, provides rare exceptions with long-lived coherence. We introduce a family of local spin-$1/2$ chains with a structured subspace that hosts a much broader range of nonthermal phenomena, such as scars, chirally propagating quasiparticles or approximate topological edge modes. These nonthermal phenomena occur at high energy densities and are protected by destructive interference that suppresses leakage into the thermal bulk. We develop a quantitative leakage theory predicting which states retain coherence and propose ways to improve the stability by inducing fast oscillations in the complement subspace. We focus on experimentally accessible models for digital quantum simulators and Rydberg atom arrays, with relevant initial states whose low entanglement is independent of system size. Our framework connects asymptotic scars, quantum cages, and parent-Hamiltonian constructions, and shows that weak ergodicity breaking in chaotic systems extends well beyond revivals of homogeneous initial states.
\end{abstract}

\maketitle

\section{Introduction}\label{sec:introduction}

 Generic quantum many-body systems are expected to approach thermal equilibrium for physical initial states~\cite{SrednickiETH, Deutsch2018ETH, dalessio_quantum_2016}. This expectation has recently been challenged by weak ergodicity breaking, where special states evade full thermalization, usually in the form of periodic revivals, slow relaxation, or localization~\cite{Serbyn2021Review,Chandran2023Review,Moudgalya2022Review,papic_weak_2021}. A canonical example is quantum many-body scarring, brought into focus by experiments in Rydberg atom arrays~\cite{Bernien2017Rydberg,Bluvstein2021Controlling}, where certain initial states exhibit unexpectedly coherent revivals instead of rapid thermalization. More broadly, weak ergodicity breaking can reflect hidden structure in Hilbert space, such as constrained connectivity, fragmentation, or approximately decoupled subspaces~\cite{Shiraishi17, sala_ergodicity_2020, khemani_localization_2020}.

Only recently, it was shown that chaotic systems can also host chirally propagating quasiparticles~\cite{kerschbaumer_quasi-solitons_2026}. Multiple such quasiparticles can be created on top of scarred backgrounds~\cite{kerschbaumer_quantum_2024}, giving access to an exponentially large manifold of nonthermal states that can transport energy and information. These observations suggest that weak ergodicity breaking in chaotic systems is considerably richer than homogeneous scar dynamics alone. This raises a broader question: which nonthermal dynamics can be embedded in otherwise chaotic systems, what structures protect them, and how do they manifest in physical properties such as transport~\cite{LjubotinaPRX, morettini_2025, brighi_anomalous_2024}?

In this work, we identify a versatile mechanism for weak ergodicity breaking based on structured subspaces protected by destructive interference. We realize it in a family of local chaotic spin-$1/2$ Hamiltonians whose embedded subspaces reproduce coherent dynamics usually observed in non-chaotic models. We first introduce a two-local Hamiltonian, meaning that each term acts on at most two spins. We focus most of our discussion on models of this type, which are especially relevant for digital quantum simulators and Rydberg atom arrays, before extending the construction to longer blockade ranges. Although the structured subspace is coupled to the thermal bulk, we identify a mechanism that suppresses leakage out of it by destructive interference.

Our leakage theory pinpoints the regimes that support prolonged coherence, and bridges the gap between coherent single-particle and many-particle dynamics. This framework allows us to systematically engineer a broad range of phenomena: from scarring and slow relaxation to coherently propagating quasiparticles with nontrivial collisions, as well as approximate topological edge physics and Bloch oscillations. Our results relate to several other perspectives on weak ergodicity breaking, connecting structured-subspace dynamics to asymptotic scars~\cite{gotta_asymptotic_2023}, quantum cages~\cite{jonay2026cages,nicolau2026cages,tan2025cages,benami2025cages}, quasi-Nambu-Goldstone modes~\cite{ren_quasi-nambu-goldstone_2024}, and parent-Hamiltonian constructions~\cite{gioia_distinct_2025}. The combination of two-local terms and low-entangled relevant states also makes these models promising targets for digital quantum simulators. Together, these results provide an experimentally accessible testbed for nonthermal dynamics on quantum simulators.

\begin{figure*}[tb]
\centering
\includegraphics[width=1.0\textwidth]{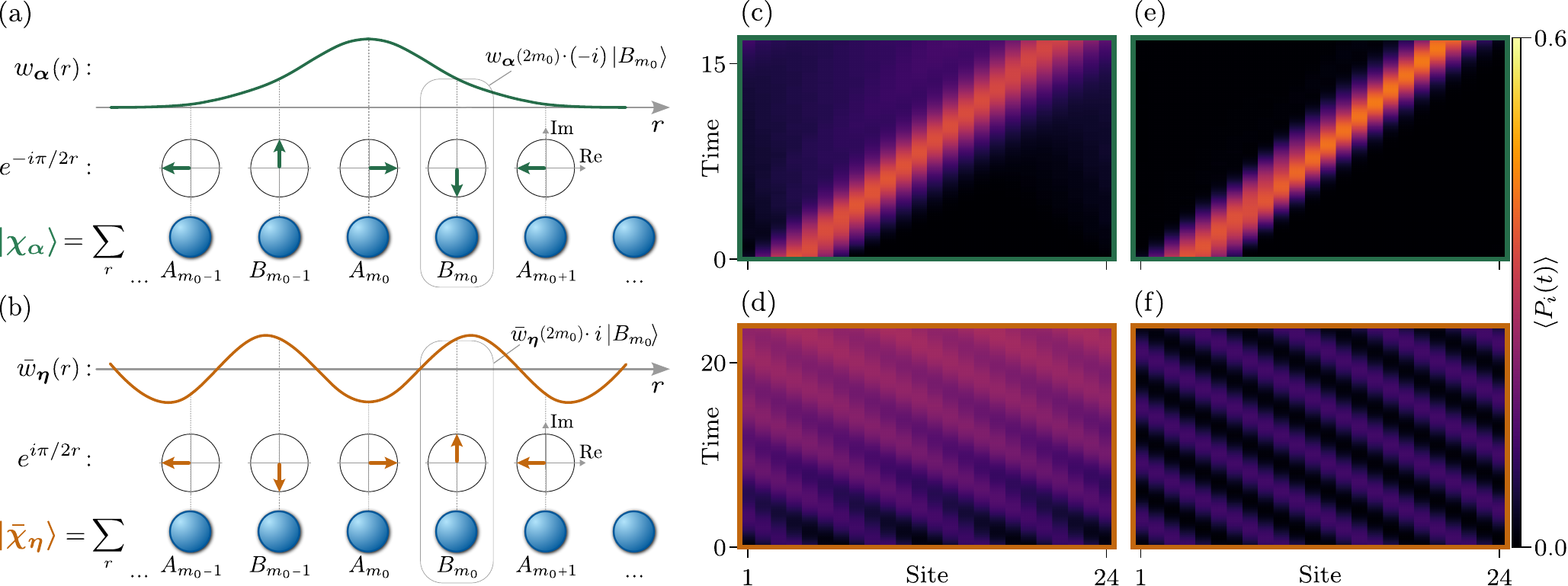}
\caption{Coherent dynamics under time-evolution of $H_\mathrm{PX}^{(g)}$ on a chain with $L=24$ sites. Schematic construction of (a) coherently propagating wavepacket $\ket{\chi_{\boldsymbol{\alpha}}}$ and (b) scarred state $\ket{\bar{\chi}_{\boldsymbol{\eta}}}$. The sketch labels $w_{\boldsymbol\alpha}$ and $\bar w_{\boldsymbol\eta}$ denote $w_{m_0,R}$ and $\bar w_{q,r_0}$, respectively. (c) Real-space dynamics up to $t=17$ for $g=2$ of $\ket{\chi_{\boldsymbol{\alpha}}}$ with $\boldsymbol{\alpha}=(7,6,-\pi/2)$. The defect propagates coherently to the right while slowly broadening in time, while the vacuum becomes only weakly perturbed. (d) Real-space dynamics for $g=2$ for the scarred state $\ket{\bar{\chi}_{\boldsymbol{\eta}}}$ with $\boldsymbol{\eta}=(2\pi/L,\pi/2)$ and envelope origin $r_0=7$ shows periodic but decaying revivals. (e) Real-space dynamics of the same wavepacket as in (c) but with $g=12$. The deformation stabilizes its coherent propagation and suppresses the spreading into the surrounding vacuum. (f) Scarred dynamics of the same state as in (d) but with $g=12$ shows almost perfect revivals.}
\label{fig:fig1_tb_dynamics}
\end{figure*}

\section{East-West model and nonthermal signatures}\label{sec:east_west}

Our starting point is the local, translation-invariant spin-$1/2$ East-West model, an inversion-symmetric relative of the kinetically constrained East model studied for its localization, nonthermal eigenstates, and slow dynamics~\cite{pancotti_quantum_2020, badbaria_state-dependent_2024}:
\begin{equation}
H_\mathrm{PX}
=
\sum_{i=1}^{L}
\big(
P_i X_{i+1}
+
X_i P_{i+1}
\big),
\label{eq:px_model}
\end{equation}
with $P_i=\ket{0}\!\bra{0}$ and $X_i=\ket{0}\!\bra{1}+\ket{1}\!\bra{0}$ the $X$ Pauli matrix. Unless noted otherwise, we use periodic boundary conditions. In comparison with the East model, the East-West model is less constrained: apart from the isolated vacuum introduced below, its computational basis states belong to a single connected sector, so there is no nontrivial Hilbert space fragmentation.
The model is chaotic, as demonstrated by symmetry-resolved spectral statistics in the Supplement~\cite{Supplement} in which we also discuss an additional symmetry present only for even system sizes.
It also anticommutes with $\mathcal C=\prod_i Z_i$, where $Z_i=\ket{0}\!\bra{0}-\ket{1}\!\bra{1}$, and its spectrum is therefore symmetric about zero energy. We discuss the spectrum and its nonthermal states in more detail below, where we also discuss an additional symmetry present only for even system sizes.

The exact eigenstate $\ket{vac}\equiv\ket{11\ldots1}$ satisfies $H_\mathrm{PX}\ket{vac}=0$ and lies at the center of the many-body spectrum, corresponding to the infinite-temperature energy density. We use it as a vacuum because it is locally annihilated by $H_\mathrm{PX}$. On top of this vacuum, we define states with single and double excitations:
\begin{equation}
\ket{A_m}=\ket{\ldots 1\,0_m\,1\ldots},
\quad
\ket{B_m}=\ket{\ldots 1\,0_m\,0_{m+1}\,1\ldots}.
\label{eq:tb_basis}
\end{equation}
Both $\ket{A_m}$ and $\ket{B_m}$ also have zero mean energy. Thus, although the vacuum is an exact eigenstate, these local defects probe the middle of the spectrum rather than low-energy excitations above a ground state. One would expect them to spread and lead the system to thermalize quickly. Surprisingly, however, excitations with long-lived coherent dynamics can be constructed from these states. We first consider wavepackets as localized excitations embedded in the vacuum of the form
\begin{equation}
\ket{\chi_{\boldsymbol{\alpha}}}
=
\sum_{r=1}^{2L}
w_{m_0,R} (r) \,e^{i k r}
\begin{cases}
\ket{A_{(r+1)/2}}, & r \text{ odd},\\[2mm]
\ket{B_{r/2}}, & r \text{ even},
\end{cases}
\label{eq:wavepacket}
\end{equation}
where $\boldsymbol{\alpha}=(m_0,R,k)$. Here $w_{m_0,R}(r)$ is a compact, approximately Gaussian envelope centered around $m_0$ with radius $R$, which localizes the defect in space; its precise form is given in the Supplement~\cite{Supplement}. The parameter $k$ sets the phase gradient of the wavepacket and resembles the momentum of the state. The construction is illustrated in Fig.~\ref{fig:fig1_tb_dynamics}(a). For $k=\pm\pi/2$, the defect propagates coherently to the left~(right) with only slow broadening, as shown in Fig.~\ref{fig:fig1_tb_dynamics}(c). It thus behaves as a chirally propagating quasiparticle embedded in an otherwise chaotic model. The stability improves as the radius $R$ is increased.

Motivated by the coherent propagation of the quasiparticle, we next consider the same Ansatz as in Eq.~\eqref{eq:wavepacket} but replace the localized envelope $w_{m_0,R}(r)$ by a spatially extended, sinusoidally modulated one. Specifically, we define $\boldsymbol{\eta}=(q,k)$ and choose the envelope function as $\bar w_{q,r_0}(r)=(1/\sqrt{L})\cos[q(r-r_0)]$, where $q$ sets the spatial period $2\pi/|q|$ in the coordinate $r$, while $r_0$ fixes the spatial offset of the modulation. The resulting state $\ket{\bar{\chi}_{\boldsymbol{\eta}}}$ is shown schematically in Fig.~\ref{fig:fig1_tb_dynamics}(b). As for the localized excitations, $k$ fixes the direction and velocity of the underlying motion. In particular, the same choice $k=\pi/2$ that supports coherent localized excitations also works for this extended state. For $q=2\pi/L$, it leads to coherent transport to the left, as shown in Fig.~\ref{fig:fig1_tb_dynamics}(d). This corresponds to scarred dynamics: the state exhibits pronounced revivals whose spatial and temporal periods are determined by $q$. In the Supplement~\cite{Supplement} we show how smaller $q$ values lead to slower oscillations and improved reviving quality while generic intial states quickly thermalize. 

Next we consider the deformation
\begin{equation}
H_\mathrm{PX}^{(g)}
=
H_\mathrm{PX}
+
(g-2)H_\mathrm{PXP},
\quad
H_\mathrm{PXP}
=
\sum_{j=1}^{L}
P_{j-1}X_jP_{j+1}.
\label{eq:hpx_g}
\end{equation}
where $g=2$ recovers the original East-West model. The added term $H_\mathrm{PXP}$ annihilates all states $\ket{A_m}$ and $\ket{B_m}$, while modifying the dynamics outside their span. The original East-West Hamiltonian in Eq.~\eqref{eq:px_model} can be decomposed as $\frac{1}{2}\sum_j \left(X_j-Z_{j-1}X_jZ_{j+1}\right)+2 H_\mathrm{PXP}$, where the first term is domain-wall conserving but the second is not. The shift by $2$ in Eq.~\eqref{eq:hpx_g} makes $g$ the coefficient of the full PXP term and controls the strength of processes that create or destroy domain-walls. For the parameters studied here, the full Hamiltonian remains chaotic, as shown in the Supplement.

We observe that the dynamics under $H_\mathrm{PX}^{(g)}$ with $g=12$ significantly stabilizes the quasiparticle. Figure~\ref{fig:fig1_tb_dynamics}(e) shows the same initial wavepacket as in panel (c), now retaining its shape while propagating through the vacuum. Likewise, the deformation stabilizes the scarred revivals in Fig.~\ref{fig:fig1_tb_dynamics}(f). Thus the same simple modification improves both the propagating and the spatially extended coherent dynamics. The mechanism behind this stabilization is explained below.

\begin{figure}[tb]
\centering
\includegraphics[width=0.97\columnwidth]{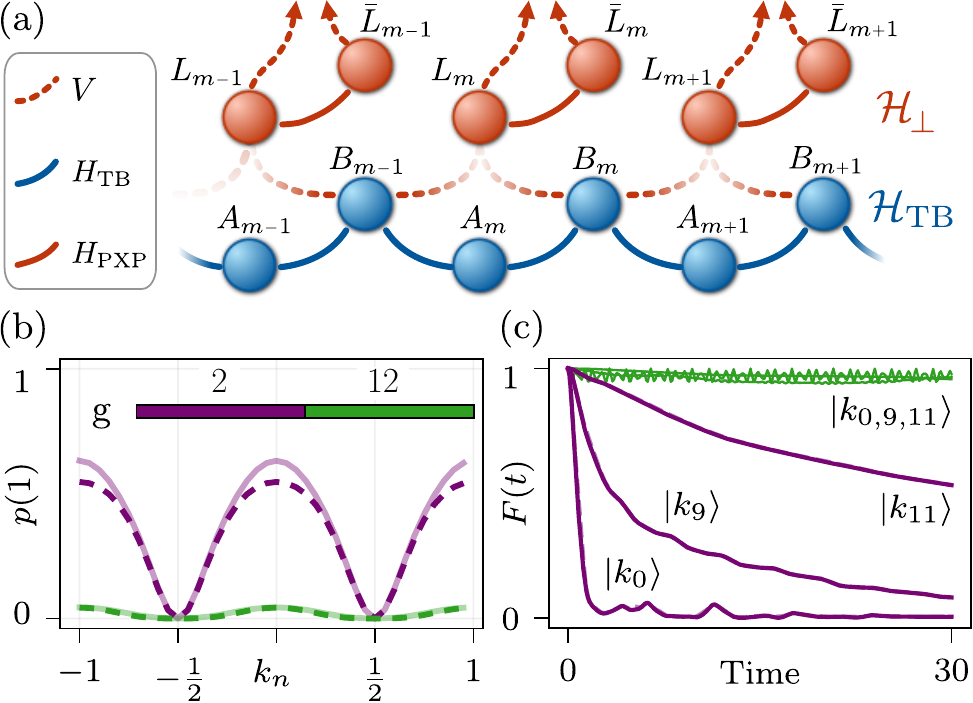}
\caption{Leakage theory for time evolution under $H_\mathrm{PX}^{(g)}$. (a) Schematic of interactions on relevant states for leakage theory. Inside $\mathcal{H}_{\mathrm{TB}}$ (blue), the dynamics is exactly tight-binding-like, while leakage originates from $V$ coupling the $\ket{B_m}$ states to $\mathcal{H}_\perp$ (red), which can cancel by destructive interference between neighboring amplitudes (visualized by fading of dashed lines). The full $gH_\mathrm{PXP}$ term acts only nontrivially in $\mathcal{H}_\perp$ and drives rapid oscillations in a small two-level subspace, thereby suppressing the accumulated leaked weight. (b) Average leakage rate between $t=0$ and $t=1$ for tight-binding momentum states $\ket{k_n}$, obtained numerically (solid) and from the analytical prediction (dashed) for $L=20$ sites. (c) Slow relaxation of different momentum states $\ket{k_n}$ close to $\pi/2$ for $g=2$ and almost no relaxation for all states for $g=12$ measured in terms of the fidelity $F(t) = |\langle k_n(0)|k_n(t) \rangle|^2$. States are defined on $L=24$ sites.}
\label{fig:fig2_leakage}%
\end{figure}

\section{Protected subspace}\label{sec:protected}
We now show how a structured subspace explains the coherent dynamics observed above. The states $\ket{A_m}$ and $\ket{B_m}$ form alternating sites of an effective tight-binding chain, while destructive interference suppresses leakage from this subspace into the rest of the Hilbert space. We quantify this suppression with a perturbative leakage theory and use the results to connect the coherent dynamics to exact and asymptotic quantum many-body scars.
\subsection{Tight-binding subspace}\label{sec:tb}

To understand the coherent dynamics observed in the East-West model and its deformed version, we use the excitations defined in Eq.~\eqref{eq:tb_basis} and define the structured subspace $\mathcal{H}_{\mathrm{TB}} = {\mathrm{span}}\{\ket{A_m},\ket{B_m}\}_{m=1}^{L}$, spanned by all single-site and nearest-neighbor two-site excitations. We define the projector onto this subspace as $\Pi_{\mathrm{TB}} = \sum_{m=1}^{L} \big(
\ket{A_m}\bra{A_m} + \ket{B_m}\bra{B_m} \big)$ and its complement $\Pi_\perp=\mathbf{1}-\Pi_{\mathrm{TB}}$. We observe that in this subspace, $H_\mathrm{PX}$ acts exactly as a single-particle tight-binding Hamiltonian $H_{\mathrm{TB}} \equiv \Pi_{\mathrm{TB}} H_\mathrm{PX}\Pi_{\mathrm{TB}}$ on a $2L$-site ring:
\begin{equation}
H_{\mathrm{TB}} = \sum_{m=1}^{L}
\Big(\ket{A_m}\bra{B_m}
+ \ket{B_m}\bra{A_{m+1}}
+ \mathrm{h.c.}\Big)
\label{eq:TB_identification}
\end{equation}
where the states $\ket{A_m}$/$\ket{B_m}$ from~Eq.~\eqref{eq:tb_basis} are identified as odd/even tight-binding sites. 

We can understand the origin of Eq.~\eqref{eq:wavepacket}: within the tight-binding subspace, it is a localized wavepacket centered around momentum $k$. The corresponding tight-binding momentum eigenstates are
\begin{equation}
\ket{k_n}
=
\frac{1}{\sqrt{2L}}
\sum_{r=1}^{2L}
e^{ik_n r}
\begin{cases}
\ket{A_{(r+1)/2}}, & r \text{ odd},\\[2mm]
\ket{B_{r/2}}, & r \text{ even},
\end{cases}
\label{eq:momentum_states}
\end{equation}
$k_n=\pi n/L$ and $n=-L,\ldots,L-1$. As in the tight-binding model, these states satisfy $H_{\mathrm{TB}}\ket{k_n}=E(k_n)\ket{k_n}$ with $E(k_n)=2\cos k_n$. The group velocity is $v_g(k_n)=\partial_k E(k_n)=-2\sin k_n$, and the dispersion curvature vanishes at $k_n=\pm\pi/2$, so wavepackets centered around these momenta broaden only weakly in the tight-binding description while propagating at maximal speed, consistent with Fig.~\ref{fig:fig1_tb_dynamics}(c). Therefore, the spatially extended, scarred states can be interpreted as a superposition of two momentum states $\ket{\bar{\chi}_{\boldsymbol{\eta}}} = (e^{-iqr_0}\ket{k_{n_1}} + e^{iqr_0}\ket{k_{n_2}})/\sqrt{2}$, with $k = (k_{n_1}+k_{n_2})/2$ and $q = (k_{n_1}-k_{n_2})/2$, where the state in Fig.~\ref{fig:fig1_tb_dynamics}(d) corresponds to $k_{n_1/n_2}=\pi/2 \pm 2\pi/L$. The reviving frequency of $\ket{\bar{\chi}_{\boldsymbol{\eta}}}$ is consequently given by $\omega = 4|\sin k\sin q|$ which is consistent with the observed dynamics in Fig.~\ref{fig:fig1_tb_dynamics}(d).

We can also understand the reduced leakage near $k=\pm\pi/2$ qualitatively from the coupling to the rest of the Hilbert space. The remaining action $H_\mathrm{PX}-H_\mathrm{TB}$ contains both dynamics within the complement $\mathcal{H}_{\perp}$ and its coupling to $\mathcal{H}_{\mathrm{TB}}$, and annihilates every $\ket{A_m}$. However, it couples to the first leaked states $\ket{L_m}=\ket{\ldots1\,0_{m-1}\,0_m\,0_{m+1}\,1\ldots}$, which are the only computational basis states directly coupled to the tight-binding subspace, when acting on $\ket{B_{m-1}}$ and $\ket{B_m}$, see Fig.~\ref{fig:fig2_leakage}(a). In a momentum state, these neighboring amplitudes differ by $e^{2ik}$ and cancel exactly at $k=\pm\pi/2$. A narrow momentum distribution around these values only weakly violates this cancellation. Together with the weak dispersion, this explains the stability of the corresponding wavepackets.

We can also intuitively understand why the deformation in Eq.~\eqref{eq:hpx_g} stabilizes this dynamics. Since $H_\mathrm{PXP}$ annihilates every $\ket{A_m}$ and $\ket{B_m}$, it leaves the projected tight-binding dynamics untouched. On a first leaked state $\ket{L_m}$, it only generates $\ket{\bar L_m}=\ket{\ldots1\,0_{m-1}\,1_m\,0_{m+1}\,1\ldots}$ and maps this state back to $\ket{L_m}$ instead of exploring the full Hilbert space. The resulting fast oscillations within a local two-level system suppress the accumulated leakage, as quantified in Sec.~\ref{sec:leakage_derivation}.

\subsection{Perturbative leakage theory}\label{sec:leakage_derivation}

To quantify the effect of destructive interference, we split the Hamiltonian as $H_\mathrm{PX}^{(g)}=H_0+V$. We choose $H_0=H_{\mathrm{TB}}+gH_\mathrm{PXP}$ as the part whose evolution we treat exactly. The remainder $V=H_\mathrm{PX}^{(g)}-H_0$ contains both dynamics within the complement of $\mathcal{H}_{\mathrm{TB}}$ and the coupling that allows states to leak into it. For an initial state $\ket{\psi_0}\in\mathcal{H}_{\mathrm{TB}}$, we go to the interaction picture with respect to $H_0$ and perform a Dyson expansion~\cite{sakurai_modern_2020} in $V$ up to first order. Including the full $H_\mathrm{PXP}$ contribution in this \emph{free part} also treats the fast oscillations of the first leaked states exactly. As shown in Fig.~\ref{fig:fig2_leakage}(a), $H_\mathrm{PXP}$ interchanges $\ket{L_m}$ and $\ket{\bar L_m}$ without coupling this pair to the rest of $\mathcal{H}_{\perp}$. Moreover, $[H_\mathrm{TB},H_\mathrm{PXP}]=0$, so the free evolution is explicit both inside the tight-binding subspace and within these two-level systems.

The Dyson expansion in $V$ does not require the full operator norm of $V$ to be small: for these initial states, its leading correction is determined by the integrated leakage out of $\mathcal H_{\mathrm{TB}}$. To first order, the interaction-picture propagator is $U_I(\tau)\approx\mathbf{1}-i\int_0^\tau ds\,V_I(s)$, with $V_I(s)=e^{iH_0s}Ve^{-iH_0s}$. Transforming back includes the free evolution generated by $H_0$. Since $H_\mathrm{PXP}$ annihilates the initial tight-binding state, the time-evolved state up to first order is $\ket{\psi(\tau)}\approx e^{-iH_{\mathrm{TB}}\tau}\ket{\psi_0}+\ket{\psi^{(1)}(\tau)}$, and the corresponding leaked component is
\begin{multline}
\ket{\psi^{(1)}(\tau)}
=
-i\int_0^\tau \!\!\!\!{\rm d}s\,
e^{-igH_\mathrm{PXP}(\tau-s)}
V
e^{-iH_{\mathrm{TB}}s}\ket{\psi_0}\\
\label{eq:first_order_dyson}
\!=\!
-i
\!\sum_{m=1}^{L}
\int_0^\tau \!\!\!\!{\rm d}s\,
e^{-igH_\mathrm{PXP}(\tau-s)}
\big[
\beta_{m-1}(s)+\beta_m(s)
\big]
\ket{L_m}\! ,
\end{multline}
where $\beta_m(s)=\langle B_m|e^{-iH_{\mathrm{TB}}s}|\psi_0\rangle$ is the amplitude of the time-evolved state $\ket{B_m}$. This equation quantifies the mechanisms identified above throughout the free evolution, as visualized in Fig.~\ref{fig:fig2_leakage}(a). First, the leakage out of the tight-binding manifold arises only from the $\ket{B_m}$ sector via the coupling of $V$. Thus the cancellation must hold along the tight-binding trajectory, rather than only in the initial state: neighboring contributions $\beta_{m-1}(s)$ and $\beta_m(s)$ can \emph{destructively interfere} at each time, as indicated by the dashed, fading lines leading to the leaked states in Fig.~\ref{fig:fig2_leakage}(a). This occurs for trajectories with momenta close to $k_n = \pm \pi/2$, because in momentum space $\beta_{m-1}(s)+\beta_m(s)=\sqrt{2/L}\sum_k e^{i(2m-1)k-i2s\cos k}\psi_k(0)\cos k$. Second, we see that the terms that do leak outside the tight-binding manifold get subsequently rotated by $H_\mathrm{PXP}$ before being accumulated in time. Consequently, for sufficiently large $g$, $H_\mathrm{PXP}$ generates fast \emph{oscillations} between $\ket{L_m}$ and $\ket{\bar L_m}$ (connected by solid, red line in Fig.~\ref{fig:fig2_leakage}(a)), so that leaked components average out and the time-integrated leaked weight is suppressed. The final expression for the leakage out of the tight-binding subspace becomes:

\begin{align}
p(\tau)
&=
\frac{\|\Pi_\perp\ket{\psi(\tau)}\|^2}
     {\|\ket{\psi(\tau)}\|^2}
\approx
\frac{\Lambda(\tau)}{1+\Lambda(\tau)},
\label{eq:pleak}
\\
\Lambda(\tau)
&=
\sum_{\substack{0\leq k_n \\ s=\pm}}
\cos^2 k_n\!
\left|
\sum_{u=0}^{1}\!
\psi_{k_n+u\pi}(0)\,
F_s(E(k_n+u\pi),g\tau)
\right|^2,
\notag
\end{align}
where the outer sum is over non-negative discrete momenta and $k_n=0$, with
$\psi_{k_n}(0)=\langle k_n|\psi_0\rangle$ being the wavefunction amplitudes in momentum basis, and
\begin{equation}
F_\pm(E,g\tau)
=
2i\,e^{-i(E\pm g)\tau/2}\,
\frac{\sin\!\big((E\pm g)\tau/2\big)}{E\pm g}.
\label{eq:pleakF}
\end{equation}
Equations~\eqref{eq:pleak} and~\eqref{eq:pleakF} make the two leakage suppression mechanisms apparent. The expression for $\Lambda(\tau)$ in Eq.~\eqref{eq:pleak} singles out momenta $k_n$ near $\pm\pi/2$, where $\cos^2 k_n$ vanishes, leading to low leakage around these points. Secondly, the term $F_\pm(E,g\tau)$ scales as $1/g$ for large $g$,
leading to a $1/g^2$ suppression in leakage. In Fig.~\ref{fig:fig2_leakage}(b), it is shown that this analytical prediction (dashed lines) is in good agreement with actual numerics (solid lines) up to the shown times of order $O(1)$. These leakage suppression mechanisms also manifest in slow relaxation of momentum states close to $\pm\pi/2$ for $g=2$ and to very slow relaxation for large $g$ across all momentum states as shown in Fig.~\ref{fig:fig2_leakage}(c).

The shown leakage interval $0\leq\tau\leq1$ in Fig.~\ref{fig:fig2_leakage}(b) resolves a natural local propagation time: a wavepacket near $k=\pm\pi/2$ moves by approximately two tight-binding sites, or one physical site, in one time unit. Because the Hamiltonian is translationally invariant, this leakage suppression does not depend on the wavepacket's position. For a scarred state, linearity gives the first-order leaked amplitude as the coherent sum of the leaked amplitudes of its two constituent momentum states. Each momentum component contributes a source proportional to $\cos k$, multiplied by its phase $e^{-iE(k)s}$. The phase changes during the revival, but the small factor $\cos k$ remains unchanged. Therefore, our analytical leakage formula is a useful diagnostic for the observed coherent dynamics beyond the shown evolution time of order unity.

\subsection{(Asymptotic) quantum many-body scars}\label{sec:spectrum}
For even system size, the cancellation at $k_n=\pm\pi/2$ is exact: $V\ket{k_n}=0$. Since these momentum states are eigenstates of $H_\mathrm{TB}$ and are annihilated by $H_\mathrm{PXP}$, they are also exact eigenstates of the full Hamiltonian $H_\mathrm{PX}^{(g)}$, and hence exact quantum many-body scars (QMBSs). Appropriate superpositions of these QMBSs in the East-West model were previously identified as Fock-space cages in Ref.~\cite{jonay2026cages}.

We next consider nearby momenta. For $g=2$, meaning $H_\mathrm{PX}^{(2)}=H_\mathrm{PX}$, and $k_n=\pm\pi/2\pm \pi d/L$ with fixed $d$ and $L$ large, Eq.~\eqref{eq:pleak} becomes to leading order:
\begin{equation}
p(\tau) \approx 2\pi^2 \sin^2\tau\, (d/L)^2.
\label{eq:momentum_leakage_scaling}
\end{equation}
Thus leakage vanishes as $d/L\to0$. The number of weakly leaking momentum states within a fixed narrow momentum window around these points scales as $O(L)$. The energy variance provides an alternative view of their slow relaxation. Since $\langle H_\mathrm{PX}\rangle_{k_n}=E(k_n)$, it is given by the squared norm of the leakage source and has the same vanishing factor:
\begin{equation}
\mathrm{Var}_{k_n}(H_\mathrm{PX})
=
\|\Pi_\perp H_\mathrm{PX}\ket{k_n}\|^2
\approx
2\pi^2(d/L)^2.
\label{eq:momentum_variance}
\end{equation}
This variance is the coefficient of $\tau^2$ in the short-time limit of the leakage. Together with their low entanglement, its vanishing as $d/L\to0$ identifies these momentum states as asymptotic QMBSs~\cite{gotta_asymptotic_2023}.

Alternatively, the momentum states can be written by application of local magnon operators on the vacuum:
\begin{equation}
\ket{k_n}
{=}
\frac{1}{\sqrt{2L}}
\sum_{j=1}^L
e^{i2k_n j}
\left(
e^{-ik_n}\sigma_j^- + \sigma_j^-\sigma_{j+1}^-
\right)
\ket{vac}.
\label{eq:momentum_operator_form}
\end{equation}
where $\sigma_j^-=\ket{0}\bra{1}$. Identifying momentum states close to $\pm\pi/2$ as asymptotic QMBSs then connects the chiral quasiparticles to the framework of Ref.~\cite{ren_quasi-nambu-goldstone_2024}, where chiral dynamics is constructed from momentum-deformed magnon operators in the presence of a quasi-symmetry. The modified magnon operators in Eq.~\eqref{eq:momentum_operator_form} allow us to use this construction to create chirally propagating defects from momentum states near the exact QMBSs $\ket{\pm\pi/2}$. In comparison, our construction does not rely on a quasi-symmetry~\cite{Ren2021Quasisymmetry}, which raises the question of which ingredients are essential for their mechanism and how it may be extended by relaxing some of its requirements.

\begin{figure}[tb]
\centering
\includegraphics[width=1.0\columnwidth]{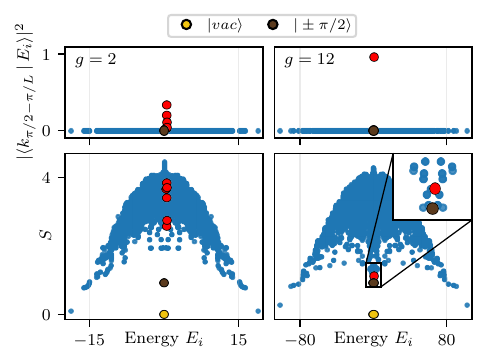}
\caption{Spectral properties of $H_\mathrm{PX}^g$ with $g=2$ (left) and $g=12$ (right) in regard to the momentum state $\ket{\pi/2-\pi/L}$ for a system with $L=14$ sites. The top row shows the overlap of the momentum state with all eigenstates of the Hamiltonians. The bottom row shows the entanglement of the spectra, where the eigenstates with high overlap with the momentum state are highlighted in red.
}
\label{fig:EM_Spectrum}%
\end{figure}

To relate these dynamical signatures to the many-body spectrum, in Fig.~\ref{fig:EM_Spectrum} we analyze the spectral decomposition of the momentum state $\ket{\pi/2-\pi/L}$, close to the exact QMBS $\ket{\pi/2}$, in the eigenbasis of $H_\mathrm{PX}^g$, quantified by its overlaps with the exact eigenstates. For $g=2$, this state has large overlap with only a few eigenstates within a narrow energy window around $E=0$, where $E=0$ is the exact eigenvalue of the QMBS $\ket{\pi/2}$, as expected for an asymptotic QMBS.
For momentum states further away from $\pm \pi/2$ this behavior is less pronounced and the states show overlap with eigenstates more evenly spread across the spectrum. Several other eigenstates with unusually low entanglement can also be seen.
For $g=12$ in comparison, the state $\ket{\pi/2-\pi/L}$ is almost a perfect eigenstate,  as it has an overlap of almost one with an eigenstate. That eigenstate also has very low entanglement, close to that of $\ket{\pi/2}$. There are many other QMBSs with similar low entanglement, as shown in the bottom-right panel. Each of these eigenstates has an overlap close to one with a momentum state. Hence, we interpret the leakage suppression term to push momentum states closer to exact QMBSs by increasing their overlap with a single eigenstate.

\section{Quasiparticle interactions and generalizations}\label{sec:extensions}
Having established the dynamics of individual quasiparticles and its protection, we turn to their interactions and to extensions of the construction. We first study collisions and late-time relaxation at finite quasiparticle density. We then realize edge states and Bloch oscillations through deformations of the East-West model and extend the construction to longer blockade ranges.
\subsection{Many quasiparticles and late time relaxation}\label{sec:finite_density}
\begin{figure}[tb]
\centering
\includegraphics[width=1.0\columnwidth]{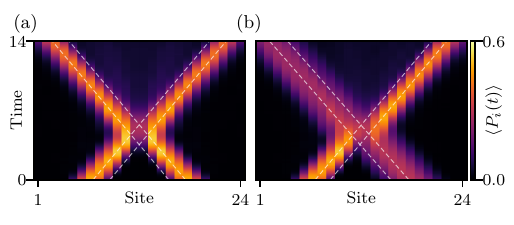}
\caption{Collisions of quasiparticles under $H_\mathrm{PX}^{(g)}$ with $g=12$ on a periodic chain with $L=24$ sites. The local down-projector density $\langle P_i(t)\rangle$ is shown for two-packet initial states $\ket{\chi_{\boldsymbol\alpha,\boldsymbol\beta}}$. (a) Counterpropagating packets with equal radii, $\boldsymbol\alpha=(13,4,-\pi/2)$ and $\boldsymbol\beta=(35,4,\pi/2)$. (b) Packets with different radii, $\boldsymbol\alpha=(13,4,-\pi/2)$ and $\boldsymbol\beta=(35,8,\pi/2)$. The quasiparticles interact during the collision but subsequently recover approximately their original form, with a phase shift relative to their incoming trajectories. Dashed lines indicate fitted trajectories; both panels use the same density scale.}
\label{fig:collisions}
\end{figure}

So far, we have considered a single-particle tight-binding subspace. Several quasiparticles can be prepared by embedding localized wavepackets on disjoint supports in the vacuum, but one would expect them to enhance leakage and substantially accelerate relaxation towards equilibrium upon collision. However, the stabilization term also allows coherent dynamics starting from such states and leads to nontrivial interactions between the quasiparticles. We denote a state with two initial wavepackets by $\ket{\chi_{\boldsymbol\alpha,\boldsymbol\beta}}$, constructed as the tensor product of the two localized wavepackets on their respective supports and $\ket{1}$ elsewhere. Figure~\ref{fig:collisions} shows collisions between counterpropagating quasiparticles for $g=12$, both with equal size and with different sizes. After colliding, the quasiparticles approximately recover their original form and acquire a phase shift. This behavior is reminiscent of solitons~\cite{dauxois_physics_2010}, although a precise classification of the quasiparticles is left for future work.

Our perturbative leakage theory identifies trajectories that leave the single-particle tight-binding subspace slowly and retain coherence for unusually long times. It does not account for the collisions just considered, and higher orders in the Dyson expansion become important once an appreciable weight has escaped. To examine the dynamics beyond this regime, we study the late-time relaxation of several quasiparticles at nonzero density numerically.

\begin{figure}[tb]
\centering
\includegraphics[width=1.0\columnwidth]{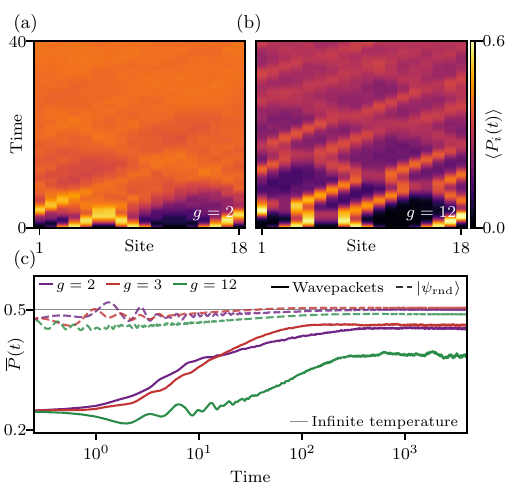}
\caption{Relaxation at finite quasiparticle density under $H_\mathrm{PX}^{(g)}$ on a periodic chain with $L=18$ sites. The initial state $\ket{\chi_{\boldsymbol\alpha,\boldsymbol\beta,\boldsymbol\gamma}}$ contains three compact packets with $\boldsymbol\alpha=(6,3,-\pi/2)$, $\boldsymbol\beta=(18,3,\pi/2)$, and $\boldsymbol\gamma=(30,3,-\pi/2)$, using the same envelope as the other localized states. (a),(b) Local down-projector density $\langle P_i(t)\rangle$ for $g=2$ and $g=12$ up to $t=40$. (c) Mean down-projector density $\overline P(t)$ for the three-packet state (solid lines) and a random product state (dashed lines), for $g=2,3,12$ up to $t=4000$. The horizontal gray line marks the infinite-temperature value $1/2$, used as an approximate reference. The random state is specified in the Supplement.}
\label{fig:finite_density}
\end{figure}

We initialize three separated quasiparticles with carrier momenta $k=\pm\pi/2$, denoted by $\ket{\chi_{\boldsymbol\alpha,\boldsymbol\beta,\boldsymbol\gamma}}$, analogously to the two-packet states in Fig.~\ref{fig:collisions} which leads to repeated collisions between the wavepackets due to the finite system size and periodic boundary conditions. The dynamics in Figs.~\ref{fig:finite_density}(a) and (b) show that the wavepackets deteriorate upon collision and spread across the system for $g=2$, while signatures of coherent propagation remain visible for $g=12$ even upon multiple collisions.

To follow the relaxation to longer times, Fig.~\ref{fig:finite_density}(c) shows the mean down-projector density, $\overline P(t)=L^{-1}\sum_i\langle P_i(t)\rangle$, for $g=2,3,12$ up to $t=4000$, compared with a random product state. At high temperature, we expect this quantity to be close to $1/2$, which serves as an approximate reference for both initial states. Within the first $40$ time units shown in panels (a) and (b), the unstabilized wavepackets have lost their visible structure and spread almost uniformly through the chain, yet their mean density continues to relax over much longer times. The state with embedded quasiparticles retains appreciable deviations from this reference value at late times, most strongly for $g=12$. The random product state approaches the reference more closely, although smaller residual offsets remain. Overlap with long-lived states containing several quasiparticles may contribute to the larger offset at $g=12$. 

Consequently, the studied quasiparticle states fail to fully thermalize even at late times for the considered system sizes. A comparison of the dynamics with smaller system with $L=12$ sites in the Supplement~\cite{Supplement} shows little size dependence of the late-time mean density for the random product state at $g=2,3$, whereas appreciable differences remain at $g=12$ and for the wavepacket states, so convergence of the $L=18$ results with system size remains unknown. Establishing whether these deviations from the thermal values persist in the thermodynamic limit requires a study of larger systems.

\subsection{Edge states and Bloch oscillations}\label{sec:edge_bloch}
\begin{figure}[tb]
\centering
\includegraphics[width=1.0\columnwidth]{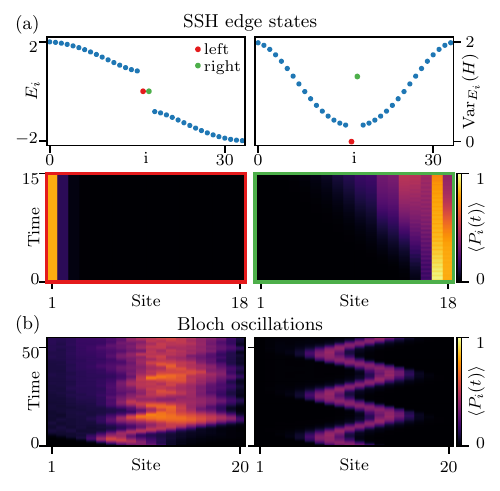}
\caption{
(a) SSH-like edge physics in the dimerized $H_{\mathrm{PX,SSH}}$ spin chain with $L=18$ and $t_1=0.6$,$t_2=1.4$. The top row shows the energy $\langle H_{\mathrm{PX,SSH}}\rangle$ and variance $\mathrm{Var}(H_{\mathrm{PX,SSH}})$ with regard to the eigenstates of $\Pi_{\mathrm{SSH}}H_{\mathrm{PX,SSH}}\Pi_{\mathrm{SSH}}$ with nonzero eigenvalues. The middle row shows the dynamics of the edge modes (given in Eq.~\eqref{eq:SSH_edge_AB}) under $H_{\mathrm{PX,SSH}}$ with $g=2$ (left) and $g=12$ (right). (b) Dynamics of $\ket{\chi_{\boldsymbol{\alpha}}}$ with $\boldsymbol{\alpha}=(21,6,\pi/2)$ defined on $L=20$ sites under evolution with $H_{\mathrm{Bloch}}^g$ with $F=0.6$. The left plot shows the evolution with $g=2$ which leads only to initial signatures of the Bloch oscillations until leakage becomes too large, while the right plots with $g=12$ shows clean and persistent Bloch oscillations.}
\label{fig:fig4_spt_bloch}%
\end{figure}

Having realized coherent propagation and collisions, we now show how changing the interactions while retaining the same defect basis gives access to other single-particle phenomena. An interesting example is the Su-Schrieffer-Heeger (SSH) model~\cite{mondal_su-schrieffer-heeger_2025}, a tight-binding model with alternating strong and weak hopping amplitudes. With weak bonds at its boundaries, it realizes a nontrivial topological phase supporting localized edge states. To reproduce this structure in our subspace, we introduce modulated ``hopping'', set by the spin coupling strengths, and use open boundary conditions:
\begin{equation}
H_{\mathrm{PX,SSH}}
=
\sum_{i=1}^{L-2}
\big(
t_1 P_iX_{i+1}
+
t_2 X_iP_{i+1}
\big)
+t_1 P_{L-1}X_L.
\label{eq:hpx_ssh_g}
\end{equation}
The deformed model is introduced as before, $H_{\mathrm{PX,SSH}}^{(g)} = H_{\mathrm{PX,SSH}}+(g-2)\sum_{i=2}^{L-1} P_{i-1}X_iP_{i+1}$.  The Hamiltonian $H_{\mathrm{PX,SSH}}$ acts as the SSH model in the non-trivial topological phase when $|t_2|>|t_1|$ within the open-chain subspace defined as
$\mathcal{H}_{\mathrm{SSH}}=\mathrm{span}\{\ket{A_1},\ket{B_1},\ldots,\ket{A_{L-1}},\ket{B_{L-1}}\}$, with $\Pi_{\mathrm{SSH}}$ being the projector onto this subspace. The boundary term in Eq.~\eqref{eq:hpx_ssh_g} removes the link to $\ket{A_L}$, so the projected model has $2(L-1)$ sites with alternating couplings $t_1$ and $t_2$ and weak bonds at both ends. The stabilization term annihilates this subspace and only suppresses leakage out of it.

To verify that the structured subspace has the expected SSH spectral properties, we diagonalize $\Pi_{\mathrm{SSH}}H_{\mathrm{PX,SSH}}\Pi_{\mathrm{SSH}}$ and obtain SSH-like eigenpairs
$\{E_i,\ket{E_i}\}$, with energies $E_i$ shown in Fig.~\ref{fig:fig4_spt_bloch}(a), including the characteristic edge states within the bulk gap. Although $E_i=\langle E_i|H_{\mathrm{PX,SSH}}|E_i\rangle$, the states $\ket{E_i}$ are not exact eigenstates of the full Hamiltonian $H_{\mathrm{PX,SSH}}$ because of leakage. To quantify how close they are to true eigenstates, we compute the energy variance
$\mathrm{Var}_{E_i}(H_{PX,\mathrm{SSH}})=
\langle E_i|H_{\mathrm{PX,SSH}}^2| E_i\rangle
-\langle E_i|H_{\mathrm{PX,SSH}}| E_i\rangle^2$.

The localized left and right edge states used in Fig.~\ref{fig:fig4_spt_bloch}(a) are
\begin{equation}
|\psi_l\rangle \propto \sum_{m=1}^{L-1}\hspace{-0.12cm}\left(-\frac{t_1}{t_2}\right)^{m-1}\hspace{-0.55cm}|A_m\rangle,\ \ |\psi_r\rangle \propto \sum_{m=1}^{L-1}\hspace{-0.12cm}\left(-\frac{t_1}{t_2}\right)^{L-1-m}\hspace{-0.85cm}|B_m\rangle.
\label{eq:SSH_edge_AB}
\end{equation}
The left state is supported entirely on the $\ket{A_m}$ sublattice, which does not couple out of the structured subspace. For a finite system, it has an exponentially small residual at the opposite edge, so its variance is exponentially small. It therefore remains localized on the displayed time scale. By contrast, the right edge state lives on the $\ket{B_m}$ sublattice and leaks under $g=2$. Turning on the suppression term restores its stability and yields long-lived edge localization~\cite{logaric_dynamical_2026}. Thus, our approach produces approximate topological edge physics inside a chaotic spin chain.

Finally, we return to uniform East-West couplings with open boundary conditions and add a linear potential to generate Bloch oscillations. In the tight-binding description, this corresponds to linearly increasing energy with site number $n$, $\delta H=F_{\mathrm{TB}}\sum_n n\ket{n}\bra{n}$. With this deformation, a wavepacket with central momentum $k$ undergoes periodic motion with $k(t)=k-F_{\mathrm{TB}}t$ and Bloch period $T_B=2\pi/|F_{\mathrm{TB}}|$. Mapping this term back to the spin chain gives
\begin{equation}
\label{eq:tb_bloch_model}
\delta H_{\mathrm{Bloch}}=F\sum_m \big(mP_m-(m+\tfrac{1}{2})P_mP_{m+1}\big).
\end{equation}
Here $F_{\mathrm{TB}}=F/2$, since adjacent tight-binding states are separated by half a physical lattice spacing. Figure~\ref{fig:fig4_spt_bloch}(b) shows that without leakage suppression only initial signatures of Bloch oscillations survive, whereas for $g=12$ the oscillations remain coherent for long times. Together, these examples show that interference-protected subspaces can provide a flexible route to distinct nonthermal dynamical signatures.

\subsection{Longer blockade ranges}\label{sec:long_range}%
We have focused on two-local models because of their experimental relevance. The two-local East-West Hamiltonian can be generalized to a family of models with arbitrary blockade range by replacing the single projector next to the spin flip by a string of $\alpha$ projectors,
\begin{equation}
H_{\mathrm{PX}}^{(\alpha)}
=\sum_i
\left(
P_{i-\alpha+1}\cdots P_i X_{i+1} +
X_i P_{i+1}\cdots P_{i+\alpha}
\right),
\label{eq:long_range_HPX}
\end{equation}
where $\alpha\ge 1$ and all indices are defined modulo $L$ for periodic boundary conditions. The case $\alpha=1$ reduces to the two-local East-West Hamiltonian studied above. It is easy to see that the tight-binding subspace of this family of models is spanned by computational basis states with  either a string of $\alpha$ zeros or a string of $\alpha+1$ zeros embedded in the vacuum:
\begin{equation}
\begin{aligned}
\ket{A_m^{(\alpha)}} &=\ket{\ldots 1\,0_m0_{m+1}\cdots 0_{m+\alpha-1}\,1\ldots},\\
\ket{B_m^{(\alpha)}} &=\ket{\ldots 1\,0_m0_{m+1}\cdots 0_{m+\alpha}\,1\ldots},
\end{aligned}
\label{eq:long_range_basis_states}
\end{equation}
yielding $\mathcal{H}_{\mathrm{TB}}^{(\alpha)}
=
\mathrm{span}\{\ket{A_m^{(\alpha)}},\ket{B_m^{(\alpha)}}\}_{m=1}^{L}$. With the projector onto this subspace $\Pi_{\mathrm{TB}}^{(\alpha)}$, we get again a tight-binding Hamiltonian of the form:
\begin{equation}
\begin{aligned}
H_{\mathrm{TB}}^{(\alpha)}
&\equiv\Pi_{\mathrm{TB}}^{(\alpha)}H_{\mathrm{PX}}^{(\alpha)}\Pi_{\mathrm{TB}}^{(\alpha)}\\
&=\sum_{m=1}^{L}\Big(
\ket{A_m^{(\alpha)}}\bra{B_m^{(\alpha)}}
+
\ket{B_m^{(\alpha)}}\bra{A_{m+1}^{(\alpha)}}
+
\mathrm{h.c.}\Big).
\end{aligned}
\label{eq:long_range_HTB}
\end{equation}
The coupling to the complement is also analogous to the East-West model. The first states outside $\mathcal{H}_{\mathrm{TB}}^{(\alpha)}$ are strings with $\alpha+2$ zeros,
$\ket{L_m^{(\alpha)}}=\ket{\ldots 1\,0_{m-1}0_{m}\cdots 0_{m+\alpha}\,1\ldots}$,
and are created by the action of the Hamiltonian on $\ket{B_{m-1}^{(\alpha)}}$ and $\ket{B_{m}^{(\alpha)}}$. Due to this analogous structure, we expect that these models are capable of producing the same tight-binding dynamics as seen from $H_{\mathrm{PX}}$. A corresponding stabilization term is $H_{\mathrm{PXP}}^{(\alpha)}=\sum_i P_{i-\alpha}\cdots P_{i-1}X_iP_{i+1}$. It annihilates the strings of $\alpha$ and $\alpha+1$ zeros spanning the tight-binding subspace. On a first leaked string of $\alpha+2$ zeros, it flips the penultimate zero and then flips it back, generating oscillations in a local two-level system. Thus the same mechanism can suppress leakage without changing the projected tight-binding dynamics. It would be interesting to compare the leakage and coherent dynamics for different blockade ranges.

\section{Experimental implementation}\label{sec:experiment}

The models considered above where each term acts only on two spins are well suited to digital quantum simulators: local Hamiltonians with small support require less gate overhead, allowing longer evolution times. The relevant initial states can also be prepared efficiently, since all states in the tight-binding subspace have low, system-size independent entanglement across all cuts into contiguous parts. For open boundary conditions, across any contiguous bipartition, a basis state in $\mathcal{H}_{\mathrm{TB}}$ can only place the excitations entirely in the left part, entirely in the right part, or straddling the cut as a local $\ket{00}$ configuration. Therefore, the Schmidt rank of any state in $\mathcal{H}_{\mathrm{TB}}$ is at most three, implying that the von Neumann entanglement entropy is bounded by $S\leq\ln3$ ($S\leq\ln4$ for periodic boundary conditions).

Rydberg arrays provide another possible realization of both the East-West model and its stabilization. We identify $\ket{0}$ as the atomic ground state and $\ket{1}$ as the Rydberg state. For strong nearest-neighbor interactions, a Rabi drive at the bare atomic excitation frequency can flip a spin only when both neighbors are unexcited, since an excited neighbor shifts the transition out of resonance. This is the usual PXP blockade mechanism~\cite{Bernien2017Rydberg,Saffman2010Rydberg}. A second Rabi drive, resonant with the bare atomic frequency shifted by the nearest-neighbor interaction, facilitates a spin flip only when exactly one neighbor is excited. Appropriately choosing the strengths of these two drives then realizes the stabilized East-West Hamiltonian in Eq.~\eqref{eq:hpx_g} with tunable $g$. This uses the same principle of selecting transitions by their interaction shifts as the multitone construction theoretically suggested in Ref.~\cite{ValenciaTortora2024Rydberg}, but with equal spacings rather than alternating bond lengths. The implementation is approximate: the drives must remain small compared with the nearest-neighbor interaction, and longer-range interaction tails remain. The effective Hamiltonian and the necessary conditions are further discussed in the Supplement.

\section{Discussion and outlook}\label{sec:discussion}

We have shown that a structured subspace embedded in a chaotic Hamiltonian can support a broad range of nonthermal phenomena because leakage out of it is suppressed by destructive interference. This goes beyond the more familiar picture of weak ergodicity breaking in terms of isolated QMBSs alone. The structured subspace can also be understood as an arrangement of exact and nearby asymptotic QMBSs into a broader manifold supporting scarred revivals, chiral transport, and other nonthermal dynamics, with appropriate superpositions of the exact states corresponding to Fock-space cages. A complementary viewpoint is provided by the local parent-Hamiltonian construction of Ref.~\cite{gioia_distinct_2025}, which relates chiral propagation to locally annihilated QMBSs. The corresponding decomposition of the East-West Hamiltonian and this connection are discussed in the Supplement~\cite{Supplement}.

The broad range of nonthermal phenomena uncovered here motivates experimental studies. The local interactions of the Hamiltonian, acting only on two neighboring sites, and the low entanglement of the relevant initial states make digital quantum simulators an attractive platform, while a dual-tone Rabi drive offers a possible realization on Rydberg atom chains. The quasiparticle collisions, which display soliton-like features~\cite{kerschbaumer_quasi-solitons_2026}, can now be followed through repeated encounters at nonzero density, as shown in Sec.~\ref{sec:finite_density}. Quantum simulators could extend this study to larger systems and other initial densities, and clarify how the observed relaxation affects transport. 

More broadly, our findings motivate the search for forms of weak ergodicity breaking beyond quantum many-body scarring of homogeneous states. Inhomogeneous quenches can reveal a rich range of dynamics, and quantum simulators can offer a way to explore such processes in regimes that are difficult to access classically. It would be particularly interesting to investigate whether our framework extends to two-dimensional systems. The extension to longer blockade ranges already suggests that interference-protected structured subspaces occur in a wider family of models. Our construction uses the computational basis; other choices of basis, including entangled states, may support further forms of coherent dynamics. Bosonic and fermionic systems also provide promising settings, where the available local states and the sign structure imposed by fermionic statistics may offer additional ways to engineer destructive interference. These possibilities motivate a systematic construction of structured subspaces protected by destructive interference. Analogous to constructions for QMBSs~\cite{mark_eta_2020, moudgalya_eta_2020, Shiraishi17} and asymptotic QMBSs~\cite{kunimi_systematic_2025}, such an approach could provide a broad route to engineering nonthermal dynamics in chaotic systems.

\begin{acknowledgments}
The authors are grateful to Zlatko Papi\'c, Dolev Bluvstein, Nishad Maskara, Marcello Dalmonte, Thomas Iadecola, Marko Ljubotina, Jie Ren, Yupeng Wang, Lincoln Carr, Xiangyu Cao, and Johannes Feldmeier for insightful discussions. This research is funded in whole or in part by the Austrian Science Fund (FWF) \href{https://doi.org/10.55776/COE1}{10.55776/COE1} and the European Union --  NextGenerationEU. J.-Y.D.~acknowledges funding from the European Union’s Horizon 2020 research and innovation programme under the Marie Sk\l odowska-Curie Grant Agreement No.~101034413.

\noindent\textbf{Note added.}
During the preparation of this manuscript, we became aware of related work by Cao~\cite{cao_quantum_2026}, which relates so-called ``growth quenches'' from locally perturbed vacua and operator growth in the Heisenberg picture, and applies Krylov methods to the same East-West model studied here, identifying Fock-space cages as conserved charges governing transport.

\noindent\textbf{Data availability.}
The data will be provided upon reasonable request.

\noindent\textbf{Declaration of interests.}
The authors declare no competing interests.
\end{acknowledgments}

\bibliography{biblio}

@article{Shiraishi17,
  title = {Systematic Construction of Counterexamples to the Eigenstate Thermalization Hypothesis},
  author = {Shiraishi, Naoto and Mori, Takashi},
  journal = {Phys. Rev. Lett.},
  volume = {119},
  issue = {3},
  pages = {030601},
  numpages = {6},
  year = {2017},
  month = {Jul},
  publisher = {American Physical Society},
  doi = {10.1103/PhysRevLett.119.030601},
}

@article{dalessio_quantum_2016,
	title = {From quantum chaos and eigenstate thermalization to statistical mechanics and thermodynamics},
	volume = {65},
	issn = {0001-8732, 1460-6976},
	url = {http://www.tandfonline.com/doi/full/10.1080/00018732.2016.1198134},
	doi = {10.1080/00018732.2016.1198134},
	number = {3},
	urldate = {2026-06-18},
	journal = {Advances in Physics},
	author = {D'Alessio, Luca and Kafri, Yariv and Polkovnikov, Anatoli and Rigol, Marcos},
	month = may,
	year = {2016},
	pages = {239--362},
}

@article{LjubotinaPRX,
  title = {Superdiffusive Energy Transport in Kinetically Constrained Models},
  author = {Ljubotina, Marko and Desaules, Jean-Yves and Serbyn, Maksym and Papi\ifmmode \acute{c}\else \'{c}\fi{}, Zlatko},
  journal = {Phys. Rev. X},
  volume = {13},
  issue = {1},
  pages = {011033},
  numpages = {14},
  year = {2023},
  month = {Mar},
  publisher = {American Physical Society},
  doi = {10.1103/PhysRevX.13.011033},
  url = {https://link.aps.org/doi/10.1103/PhysRevX.13.011033}
}

@article{Deutsch2018ETH,
doi = {10.1088/1361-6633/aac9f1},
url = {https://dx.doi.org/10.1088/1361-6633/aac9f1},
year = {2018},
month = {jul},
publisher = {IOP Publishing},
volume = {81},
number = {8},
pages = {082001},
author = {Joshua M Deutsch},
title = {Eigenstate thermalization hypothesis},
journal = {Reports on Progress in Physics},
}

@article{SrednickiETH,
	Author = {Srednicki, Mark},
	Doi = {10.1103/PhysRevE.50.888},
	Issue = {2},
	Journal = {Phys. Rev. E},
	Month = {Aug},
	Numpages = {0},
	Pages = {888},
	Publisher = {American Physical Society},
	Title = {Chaos and quantum thermalization},
	Url = {http://link.aps.org/doi/10.1103/PhysRevE.50.888},
	Volume = {50},
	Year = {1994}}

@article{Serbyn2021Review,
  title={Quantum many-body scars and weak breaking of ergodicity},
  author={Serbyn, Maksym and Abanin, Dmitry A and Papi{\'c}, Zlatko},
  journal={Nature Physics},
  volume={17},
  number={6},
  pages={675--685},
  year={2021},
  doi={https://doi.org/10.1038/s41567-021-01230-2},
  publisher={Nature Publishing Group}
}

@article{Chandran2023Review,
  title={Quantum many-body scars: A quasiparticle perspective},
  author={Chandran, Anushya and Iadecola, Thomas and Khemani, Vedika and Moessner, Roderich},
  journal={Annual Review of Condensed Matter Physics},
  volume={14},
  pages={443--469},
  year={2023},
  publisher={Annual Reviews},
  doi={https://doi.org/10.1146/annurev-conmatphys-031620-101617}
}

@article{Moudgalya2022Review,
  title={Quantum many-body scars and {Hilbert} space fragmentation: A review of exact results},
  author={Moudgalya, Sanjay and Bernevig, B Andrei and Regnault, Nicolas},
  journal={Reports on Progress in Physics},
  volume={85},
  number={8},
  pages={086501},
  year={2022},
  publisher={IOP Publishing Ltd.},
  doi={10.1088/1361-6633/ac73a0}
}

@article{Bluvstein2021Controlling,
  title={Controlling quantum many-body dynamics in driven {Rydberg} atom arrays},
  author={Bluvstein, Dolev and Omran, Ahmed and Levine, Harry and Keesling, Alexander and Semeghini, Giulia and Ebadi, Sepehr and Wang, Tout T and Michailidis, Alexios A and Maskara, Nishad and Ho, W Wei and others},
  journal={Science},
  volume={371},
  number={6536},
  pages={1355--1359},
  year={2021},
  publisher={American Association for the Advancement of Science},
  doi= {10.1126/science.abg2530}
}

@article{Ren2021Quasisymmetry,
  title = {Quasisymmetry Groups and Many-Body Scar Dynamics},
  author = {Ren, Jie and Liang, Chenguang and Fang, Chen},
  journal = {Phys. Rev. Lett.},
  volume = {126},
  issue = {12},
  pages = {120604},
  numpages = {6},
  year = {2021},
  month = {Mar},
  publisher = {American Physical Society},
  doi = {10.1103/PhysRevLett.126.120604},
  url = {https://link.aps.org/doi/10.1103/PhysRevLett.126.120604}
}

@article{Bernien2017Rydberg,
  title={Probing many-body dynamics on a 51-atom quantum simulator},
  author={Bernien, Hannes and Schwartz, Sylvain and Keesling, Alexander and Levine, Harry and Omran, Ahmed and Pichler, Hannes and Choi, Soonwon and Zibrov, Alexander S and Endres, Manuel and Greiner, Markus and others},
  journal={Nature},
  volume={551},
  number={7682},
  pages={579--584},
  year={2017},
  doi={https://doi.org/10.1038/nature24622},
  publisher={Nature Publishing Group}
}

@article{kerschbaumer_quantum_2024,
  title = {Quantum Many-Body Scars beyond the {PXP} Model in {Rydberg} Simulators},
  author = {Kerschbaumer, Aron and Ljubotina, Marko and Serbyn, Maksym and Desaules, Jean-Yves},
  journal = {Phys. Rev. Lett.},
  volume = {134},
  issue = {16},
  pages = {160401},
  numpages = {9},
  year = {2025},
  month = {Apr},
  publisher = {American Physical Society},
  doi = {10.1103/PhysRevLett.134.160401},
}

@article{morettini_2025,
  title = {Transport in a system with a tower of quantum many-body scars},
  author = {Morettini, Gianluca and Capizzi, Luca and Fagotti, Maurizio and Mazza, Leonardo},
  journal = {Phys. Rev. B},
  volume = {112},
  issue = {13},
  pages = {134314},
  numpages = {12},
  year = {2025},
  month = {Oct},
  publisher = {American Physical Society},
  doi = {10.1103/821h-8yjz},
}

@article{ren_quasi-nambu-goldstone_2024,
	title = {Quasi-{Nambu}-{Goldstone} modes in many-body scar models},
	volume = {110},
	issn = {2469-9950, 2469-9969},
	url = {https://link.aps.org/doi/10.1103/PhysRevB.110.245101},
	doi = {10.1103/PhysRevB.110.245101},
	number = {24},
	urldate = {2026-03-24},
	journal = {Physical Review B},
	author = {Ren, Jie and Wang, Yu-Peng and Fang, Chen},
	month = dec,
	year = {2024},
	pages = {245101},
}

@article{jonay2026cages,
  title = {Localized {Fock} space cages in kinetically constrained models},
  author = {Jonay, Cheryne and Pollmann, Frank},
  journal = {Phys. Rev. B},
  volume = {113},
  issue = {13},
  pages = {134313},
  numpages = {10},
  year = {2026},
  month = {Apr},
  publisher = {American Physical Society},
  doi = {10.1103/wz33-vczt},
  url = {https://link.aps.org/doi/10.1103/wz33-vczt}
}

@article{nicolau2026cages,
  title = {Fragmentation, Zero Modes, and Collective Bound States in Constrained Models},
  author = {Nicolau, Eloi and Ljubotina, Marko and Serbyn, Maksym},
  journal = {PRX Quantum},
  volume = {7},
  issue = {1},
  pages = {010352},
  numpages = {21},
  year = {2026},
  month = {Mar},
  publisher = {American Physical Society},
  doi = {10.1103/sl79-1xgb},
  url = {https://link.aps.org/doi/10.1103/sl79-1xgb}
}

@article{tan2025cages,
      title={Interference-caged quantum many-body scars: the {Fock} space topological localization and interference zeros}, 
      author={Tao-Lin Tan and Yi-Ping Huang},
      year={2025},
      eprint={2504.07780},
      archivePrefix={arXiv},
      primaryClass={cond-mat.str-el},
      journal={arXiv e-Prints},  
}

@misc{benami2025cages,
      title={Many-body cages: disorder-free glassiness from flat bands in {Fock} space, and many-body {Rabi} oscillations}, 
      author={Tom Ben-Ami and Markus Heyl and Roderich Moessner},
      year={2025},
      eprint={2504.13086},
      archivePrefix={arXiv},
      primaryClass={cond-mat.quant-gas},
      journal={arXiv e-Prints}, 
}

@article{gioia_distinct_2025,
      title={Distinct Types of Parent Hamiltonians for Quantum States: Insights from the {$W$} State as a Quantum Many-Body Scar}, 
      author={Lei Gioia and Sanjay Moudgalya and Olexei I. Motrunich},
      year={2026},
      eprint={2510.24713},
      archivePrefix={arXiv},
      primaryClass={quant-ph},
      journal={arXiv e-Prints}, 
}

@article{kerschbaumer_quasi-solitons_2026,
	title = {Quasi-solitons in {Rydberg} atom chains},
	volume = {17},
	issn = {2041-1723},
	url = {https://www.nature.com/articles/s41467-026-75598-1},
	doi = {10.1038/s41467-026-75598-1},
	number = {1},
	urldate = {2026-12-02},
	journal = {Nature Communications},
	author = {Kerschbaumer, Aron and Desaules, Jean-Yves and Ljubotina, Marko and Serbyn, Maksym},
	month = jul,
	year = {2026},
	pages = {8783},
}

@article{gotta_asymptotic_2023,
	title = {Asymptotic Quantum Many-Body Scars},
	volume = {131},
	issn = {0031-9007, 1079-7114},
	url = {https://link.aps.org/doi/10.1103/PhysRevLett.131.190401},
	doi = {10.1103/PhysRevLett.131.190401},
	number = {19},
	urldate = {2026-03-25},
	journal = {Physical Review Letters},
	author = {Gotta, Lorenzo and Moudgalya, Sanjay and Mazza, Leonardo},
	month = nov,
	year = {2023},
	pages = {190401},
}

@Inbook{papic_weak_2021,
    author="Papi{\'{c}}, Zlatko",
    editor="Bayat, Abolfazl
    and Bose, Sougato
    and Johannesson, Henrik",
    title="Weak Ergodicity Breaking Through the Lens of Quantum Entanglement",
    bookTitle="Entanglement in Spin Chains: From Theory to Quantum Technology Applications",
    year="2022",
    publisher="Springer International Publishing",
    address="Cham",
    pages="341--395",
    isbn="978-3-031-03998-0",
    doi="10.1007/978-3-031-03998-0_13",
    url="https://doi.org/10.1007/978-3-031-03998-0_13"
}

@article{brighi_anomalous_2024,
	title = {Anomalous transport in the kinetically constrained quantum {East}-{West} model},
	volume = {110},
	issn = {2469-9950, 2469-9969},
	url = {https://link.aps.org/doi/10.1103/PhysRevB.110.L100304},
	doi = {10.1103/PhysRevB.110.L100304},
	number = {10},
	urldate = {2026-04-06},
	journal = {Physical Review B},
	author = {Brighi, Pietro and Ljubotina, Marko},
	month = sep,
	year = {2024},
	pages = {L100304},
}

@article{pancotti_quantum_2020,
	title = {Quantum {East} Model: Localization, Nonthermal Eigenstates, and Slow Dynamics},
	volume = {10},
	issn = {2160-3308},
	shorttitle = {Quantum {East} {Model}},
	url = {https://link.aps.org/doi/10.1103/PhysRevX.10.021051},
	doi = {10.1103/PhysRevX.10.021051},
	number = {2},
	urldate = {2025-08-13},
	journal = {Physical Review X},
	author = {Pancotti, Nicola and Giudice, Giacomo and Cirac, J. Ignacio and Garrahan, Juan P. and Bañuls, Mari Carmen},
	month = jun,
	year = {2020},
	pages = {021051},
}

@article{badbaria_state-dependent_2024,
	title = {State-Dependent Mobility Edge in Kinetically Constrained Models},
	volume = {5},
	issn = {2691-3399},
	url = {https://link.aps.org/doi/10.1103/PRXQuantum.5.040348},
	doi = {10.1103/PRXQuantum.5.040348},
	number = {4},
	urldate = {2026-04-12},
	journal = {PRX Quantum},
	author = {Badbaria, Manthan and Pancotti, Nicola and Singh, Rajeev and Marino, Jamir and Valencia-Tortora, Riccardo J.},
	month = dec,
	year = {2024},
	pages = {040348},
}

@article{mondal_su-schrieffer-heeger_2025,
	title = {Su-{Schrieffer}-{Heeger} {Model} - {From} {Fundamentals} to {Responses}},
	volume = {64},
	issn = {1572-9575},
	url = {https://link.springer.com/10.1007/s10773-025-05981-z},
	doi = {10.1007/s10773-025-05981-z},
	number = {5},
	urldate = {2026-04-12},
	journal = {International Journal of Theoretical Physics},
	author = {Mondal, Deep and Bandyopadhyay, Arka and Jana, Debnarayan},
	month = apr,
	year = {2025},
	pages = {125},
}

@book{dauxois_physics_2010,
	address = {Cambridge},
	title = {Physics of Solitons},
	isbn = {978-0-521-14360-8},
	publisher = {Cambridge University Press},
	author = {Dauxois, Thierry and Peyrard, Michel},
	year = {2010},
}

@article{sala_ergodicity_2020,
	title = {Ergodicity {Breaking} {Arising} from {Hilbert} {Space} {Fragmentation} in {Dipole}-{Conserving} {Hamiltonians}},
	volume = {10},
	issn = {2160-3308},
	url = {https://link.aps.org/doi/10.1103/PhysRevX.10.011047},
	doi = {10.1103/PhysRevX.10.011047},
	number = {1},
	urldate = {2026-05-21},
	journal = {Physical Review X},
	author = {Sala, Pablo and Rakovszky, Tibor and Verresen, Ruben and Knap, Michael and Pollmann, Frank},
	month = feb,
	year = {2020},
	pages = {011047},
}

@article{khemani_localization_2020,
	title = {Localization from {Hilbert} space shattering: {From} theory to physical realizations},
	volume = {101},
	issn = {2469-9950, 2469-9969},
	shorttitle = {Localization from {Hilbert} space shattering},
	url = {https://link.aps.org/doi/10.1103/PhysRevB.101.174204},
	doi = {10.1103/PhysRevB.101.174204},
	number = {17},
	urldate = {2026-05-21},
	journal = {Physical Review B},
	author = {Khemani, Vedika and Hermele, Michael and Nandkishore, Rahul},
	month = may,
	year = {2020},
	pages = {174204},
}

@book{sakurai_modern_2020,
	edition = {3},
	title = {Modern {Quantum} {Mechanics}},
	copyright = {https://www.cambridge.org/core/terms},
	isbn = {978-1-108-58728-0 978-1-108-47322-4},
	url = {https://www.cambridge.org/core/product/identifier/9781108587280/type/book},
	doi = {10.1017/9781108587280},
	urldate = {2026-05-21},
	publisher = {Cambridge University Press},
	author = {Sakurai, J. J. and Napolitano, Jim},
	month = sep,
	year = {2020},
}

@misc{logaric_dynamical_2026,
	title = {Dynamical signatures of conventional and asymptotic quantum many-body scars on a trapped ion simulator},
	url = {http://arxiv.org/abs/2604.12296},
	doi = {10.48550/arXiv.2604.12296},
	urldate = {2026-06-18},
	publisher = {arXiv},
	author = {Logarić, Leonard and Goold, John and Dooley, Shane},
	month = apr,
	year = {2026},
	note = {arXiv:2604.12296 [quant-ph]},
}

@article{kunimi_systematic_2025,
	title = {Systematic construction of asymptotic quantum many-body scar states and their relation to supersymmetric quantum mechanics},
	volume = {7},
	issn = {2643-1564},
	url = {https://link.aps.org/doi/10.1103/zstl-s1y2},
	doi = {10.1103/zstl-s1y2},
	number = {4},
	urldate = {2026-06-15},
	journal = {Physical Review Research},
	author = {Kunimi, Masaya and Kato, Yusuke and Katsura, Hosho},
	month = oct,
	year = {2025},
	pages = {043107},
}

@article{mark_eta_2020,
	title = {$\eta$ - pairing states as true scars in an extended {Hubbard} model},
	volume = {102},
	issn = {2469-9950, 2469-9969},
	url = {https://link.aps.org/doi/10.1103/PhysRevB.102.075132},
	doi = {10.1103/PhysRevB.102.075132},
	number = {7},
	urldate = {2026-06-15},
	journal = {Physical Review B},
	author = {Mark, Daniel K. and Motrunich, Olexei I.},
	month = aug,
	year = {2020},
	pages = {075132},
}

@article{moudgalya_eta_2020,
	title = {$\eta$ -pairing in {Hubbard} models: {From} spectrum generating algebras to quantum many-body scars},
	volume = {102},
	issn = {2469-9950, 2469-9969},
	shorttitle = {η -pairing in {Hubbard} models},
	url = {https://link.aps.org/doi/10.1103/PhysRevB.102.085140},
	doi = {10.1103/PhysRevB.102.085140},
	number = {8},
	urldate = {2026-06-15},
	journal = {Physical Review B},
	author = {Moudgalya, Sanjay and Regnault, Nicolas and Bernevig, B. Andrei},
	month = aug,
	year = {2020},
	pages = {085140},
}

@article{Supplement,
title={ },
journal={See Supplementary Material for additional results and details},
doi={ }
}

@article{Saffman2010Rydberg,
  title = {Quantum information with Rydberg atoms},
  author = {Saffman, M. and Walker, T. G. and M{\o}lmer, K.},
  journal = {Reviews of Modern Physics},
  volume = {82},
  pages = {2313--2363},
  year = {2010},
  doi = {10.1103/RevModPhys.82.2313}
}

@article{ValenciaTortora2024Rydberg,
  title = {Rydberg Platform for Nonergodic Chiral Quantum Dynamics},
  author = {Valencia-Tortora, Riccardo J. and Pancotti, Nicola and Fleischhauer, Michael and Bernien, Hannes and Marino, Jamir},
  journal = {Physical Review Letters},
  volume = {132},
  pages = {223201},
  year = {2024},
  doi = {10.1103/PhysRevLett.132.223201}
}

@article{cao_quantum_2026,
	title = {Quantum quenches that resemble operator growth},
	volume = {114},
	issn = {2469-9950, 2469-9969},
	url = {https://link.aps.org/doi/10.1103/4k2f-xvq3},
	doi = {10.1103/4k2f-xvq3},
	number = {20},
	urldate = {2026-12-02},
	journal = {Physical Review B},
	author = {Cao, Xiangyu},
	month = oct,
	year = {2026},
	pages = {204310},
}

\clearpage
\hypersetup{pageanchor=false}
\begingroup

\onecolumngrid
\begin{center}
\textbf{\large Supplemental Online Material for ``Coherent Dynamics in Chaotic Spin Chains via Interference-Protected Subspaces" }\\[5pt]
Aron Kerschbaumer, Jean-Yves Desaules, and Maksym Serbyn \\
{\small \sl ${}^1$Institute of Science and Technology Austria (ISTA)}

\vspace{0.1cm}
\begin{quote}
{\small In this Supplementary Material, we give the compact wavepacket envelope in Section~\ref{SM:quasiparticles} and discuss scarred revivals in Section~\ref{SM:scarred_dynamics}. Section~\ref{SM:parent_hamiltonians} gives the local parent-Hamiltonian decomposition. Section~\ref{SM:spectrum} discusses the symmetries of the East-West model and demonstrates its chaotic spectral statistics. Section~\ref{SM:rydberg} derives the proposed Rydberg implementation, and Section~\ref{SM:rnd_product_state} specifies the random product state used in Fig.~5 of the main text and Fig.~\ref{fig:scarred_dynamics} below.
} \\[10pt]

\end{quote}
\end{center}
\setcounter{equation}{0}
\setcounter{figure}{6}
\setcounter{table}{0}
\setcounter{page}{1}
\setcounter{section}{0}
\setcounter{secnumdepth}{3}
\makeatletter
\renewcommand{\theequation}{S\arabic{equation}}
\renewcommand{\thefigure}{\arabic{figure}}
\renewcommand{\thesection}{S\arabic{section}}
\renewcommand{\thepage}{\arabic{page}}
\renewcommand{\thetable}{S\arabic{table}}

\vspace{0cm}

\section{Wavepackets}\label{SM:quasiparticles}
The compact envelope $w_{m_0,R}(r)$ used for the localized states in Figs.~1, 4, 5, and 6 of the main text is given by
$w_{m_0,R}(r)=\big|\cos\!\big(\pi d_{\mathrm p}(r,m_0)/(2R+4)\big)\big|
\Theta\!\big(R-d_{\mathrm p}(r,m_0)\big)/\mathcal{N}_{m_0,R}$,
where $d_{\mathrm p}(r,m_0)$ is the shortest distance between $r$ and $m_0$ on the periodic tight-binding ring, and $\Theta(x)$ is the Heaviside step function.
Thus, the profile is restricted to distances $d_{\mathrm p}(r,m_0)\le R$. The normalization constant is
$\mathcal{N}_{m_0,R}=\sqrt{R+1+\cos\!\left(2\pi/(2R+4)\right)}$.
The same envelope is used for all compact wavepackets, including the finite-density example in Fig.~5.

\section{Scarred dynamics}\label{SM:scarred_dynamics}
In the main text, we observed the scarred dynamics of $\ket{\bar\chi_{\boldsymbol\eta}}$ through local expectation values in Fig.~1. To quantify the quality of the revivals more directly, we now consider fidelity revivals of two scarred states, $\ket{\bar\chi_{\boldsymbol\eta_1}}$ and $\ket{\bar\chi_{\boldsymbol\eta_2}}$, with $\boldsymbol\eta_1=(\pi/L,\pi/2)$ and $\boldsymbol\eta_2=(3\pi/L,\pi/2)$. Both use envelope origin $r_0=1$. These states are superpositions of the tight-binding momentum states at $k=\pi/2\pm\pi/L$ and $k=\pi/2\pm3\pi/L$, respectively. Figure~\ref{fig:scarred_dynamics}(a) shows that under time evolution with the unstabilized Hamiltonian, both scarred states retain a large fraction of their weight inside the small tight-binding manifold for long times and exhibit pronounced periodic fidelity revivals. The revivals are strongest for $\ket{\bar\chi_{\boldsymbol\eta_1}}$, whose constituent momenta lie closer to $\pi/2$, while $\ket{\bar\chi_{\boldsymbol\eta_2}}$ revives faster but with smaller fidelity peaks. In contrast, the random product state rapidly loses its initial fidelity. For $g=12$, the same scarred states remain almost completely inside the tight-binding manifold and show nearly perfect revivals, as shown in Fig.~\ref{fig:scarred_dynamics}(b). Thus, the stabilization mechanism allows one to construct long-lived scarred dynamics with tunable revival periods by choosing different pairs of tight-binding momentum states.

\begin{figure}[htbp]
\centering
\includegraphics[width=4.725in]{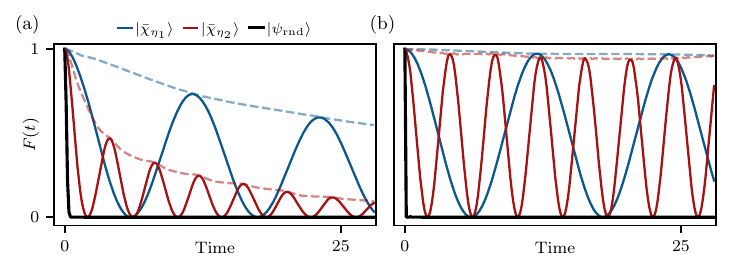}
\caption{Scarred dynamics under time evolution with $H_\mathrm{PX}^{(g)}$ on a chain with $L=24$ sites. (a) Fidelity, $F(t)=|\langle\psi(0)|\psi(t)\rangle|^2$, for $g=2$ and different initial states. The scarred state $\ket{\bar\chi_{\boldsymbol\eta_1}}$, with $\boldsymbol\eta_1=(\pi/L,\pi/2)$, is built from momentum states close to $\pi/2$ and shows pronounced revivals. The state $\ket{\bar\chi_{\boldsymbol\eta_2}}$, with $\boldsymbol\eta_2=(3\pi/L,\pi/2)$, oscillates faster but with smaller fidelity peaks. Both scarred states use envelope origin $r_0=1$. In contrast, a random product state $\ket{\psi_{\mathrm{rnd}}}$ rapidly loses its initial fidelity. Dashed lines show the weight remaining in $\mathcal H_{\mathrm{TB}}$, namely $1-p(t)$. (b) Fidelity dynamics for $g=12$ for the same initial states as in (a), showing almost perfect revivals for the scarred states.}
\label{fig:scarred_dynamics}
\end{figure}

\section{Model and symmetries}\label{SM:spectrum}
The main model we study is the (perturbed) spin-1/2 East-West quantum model
\begin{equation}
H_\mathrm{PX}^{(g)}
=\sum_{i=1}^{L}
\big(
P_i X_{i+1}
+
X_i P_{i+1}
\big)+
(g-2)\sum_{i=1}^{L}
P_{i-1}X_iP_{i+1}.
\label{eq:px_model_SM}
\end{equation}

The original East-West model corresponds to $g=2$. For any value of $g$, this model is symmetric under spatial inversion $\mathcal{I}$ and translation $\mathcal{T}$ (when periodic boundary conditions are used). It also has the chiral symmetry $\mathcal{C}=\prod_{i=1}^L Z_i$ such that $\{H_\mathrm{PX}^{(g)}, \mathcal{C}\}=0$. When $g=2$ and $L$ is even, the model has an additional symmetry $\mathcal{P}=\sum_{i=1}^L (-1)^i X_iX_{i+1}$. 
One can check that $\mathcal{P}$ indeed commutes with $H_\mathrm{PX}$ by rewriting the latter as $\sum_i P_i(X_{i-1}+X_{i+1})$ and checking that $\mathcal{P}$ commutes with each term of that sum individually as
\begin{equation}
\begin{aligned}
        [P_i(X_{i-1}+X_{i+1}),\mathcal{P}]&=(-1)^{i-1}[P_i,X_{i-1}X_{i}-X_iX_{i+1}](X_{i-1}+X_{i+1}) \\
        &=(-1)^{i-1}[P_i,X_i]X_{i-1}(X_{i-1}+X_{i+1})-(-1)^{i-1}[P_i,X_i]X_{i+1}(X_{i-1}+X_{i+1}) \\
        &=(-1)^{i-1}[P_i,X_i](1+X_{i-1}X_{i+1})-(-1)^{i-1}[P_i,X_i](X_{i-1}X_{i+1}+1)=0,
\end{aligned}
\end{equation}
where we have used the fact that $[X_i,X_j]=0$ for all $i,j$ and that $[P_i,X_j]=0$ if $i\neq j$ to take terms out of the commutator and rearrange them. We note that as the last line cancels regardless of the result of the commutator $[P_i,X_i]$.  $\mathcal{P}$ is thus a symmetry of any Hamiltonian of the form  $\sum_i h_i(X_{i-1}+X_{i+1})$, with $h_i$ an arbitrary operator that acts solely on site $i$. 

In the rest of this section, we will use $k$ to denote momentum, and $i$ and $p$ to denote the eigenvalue under $\mathcal{I}$ and $\mathcal{P}$ respectively. For even $L$, $p$ can take $2\lfloor L/4 \rfloor$ different values, from $p=-4\lfloor L/4 \rfloor$ to $p=+4\lfloor L/4 \rfloor$, each containing $\binom{L}{(L+p)/2}$ states. The sector with $p=0$ is thus always the largest. Additionally, in this sector $\mathcal{P}$ commutes with $\mathcal{T}$, which is not the case in general as for the full operator $\{\mathcal{T},\mathcal{P}\}=0$ and only $[\mathcal{T}^2,\mathcal{P}]=0$ due to the staggered structure of $\mathcal{P}$.

While we do not see any clear influence of $\mathcal{P}$ in the dynamics we study, we note that for $L$ multiple of 4 it gives rise to four exact eigenstates with $p=\pm L$. These take the form $\ket{++--++--\ldots++--}$ and its translations, with $\ket{\pm}$ denoting the eigenstate of $X$ with eigenvalue $\pm 1$. These four product states are exact zero modes. While they are not fully orthogonal to the three zero modes discussed in the main text and in Ref.~\cite{jonay2026cages}, they are linearly independent from them and orthogonal to each other. It is easy to check that they are zero modes as 
\begin{equation}\label{eq:ZE_SM_proof}
    P_i(X_{i-1}+X_{i+1})\ket{\pm}_{i-1}\ket{x}_i\ket{\mp}_{i+1}=\pm\ket{\pm}_{i-1}\left(P_i\ket{x}_i\right)\ket{\mp}_{i+1}\mp\ket{\pm}_{i-1}\left(P_i\ket{x}_i\right)\ket{\mp}_{i+1}=0,
\end{equation}
where $\ket{x}$ is an arbitrary state. As the only three-site configurations that appear in the four states of interest are $\ket{\pm}_{i-1}\ket{\pm}_i\ket{\mp}_{i+1}$ and $\ket{\pm}_{i-1}\ket{\mp}_i\ket{\mp}_{i+1}$, they are all covered by Eq.~\eqref{eq:ZE_SM_proof}. We note that this result is independent of what $\left(P_i\ket{x}_i\right)$ evaluates to and so, as for $\mathcal{P}$ being a symmetry, it must hold for any generic Hamiltonian $\sum_i h_i(X_{i-1}+X_{i+1})$. Finally, we note that as the three zero modes discussed in the main text are not eigenstates of $\mathcal{P}$, their projection into each $p$ sector is also an exact zero mode. This leads to an $\mathcal{O}(L)$ of zero modes that admit a relatively simple analytical description. 

\begin{figure*}[hbt]
\centering
\includegraphics[width=0.97\columnwidth]{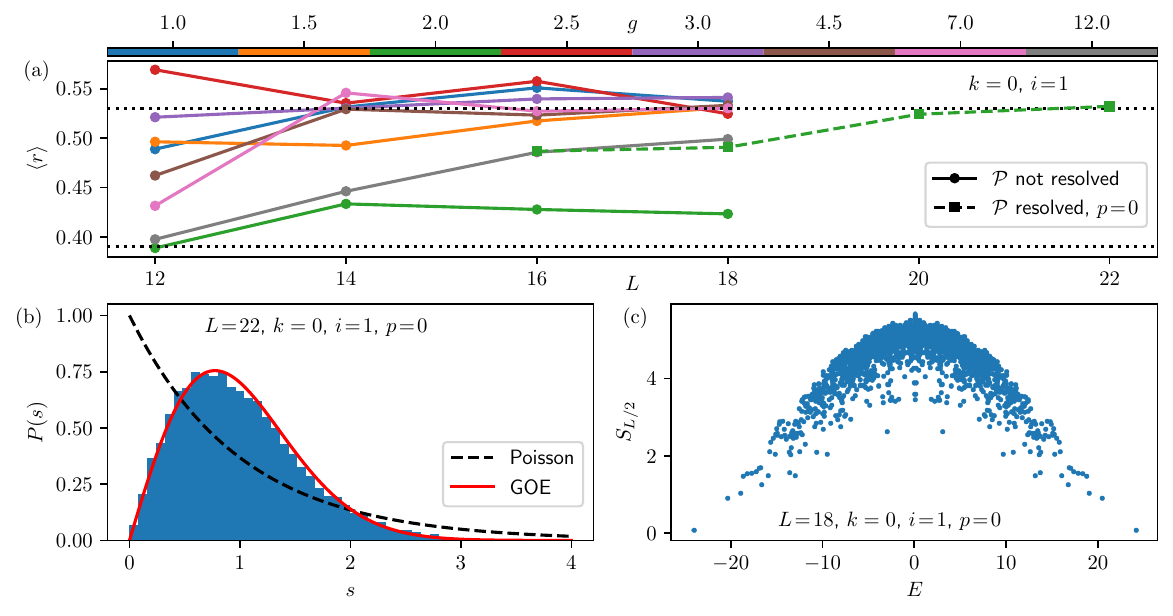}
\caption{Spectral properties of the $H_\mathrm{PX}^{(g)}$ Hamiltonian. (a) Level spacing ratio for various values of $g$ and system sizes. The two black dotted lines are at 0.39 and 0.53 and respectively denote the value for Poisson and GOE level statistics. For $g\neq2$, all data points converge towards the GOE value with increasing $L$ once $\mathcal{P}$ and $\mathcal{I}$ are resolved. For $g=2$, this convergence towards 0.53 is recovered only if $\mathcal{P}$ is resolved as well. (b) Level spacing statistics after unfolding in the most symmetric sector for $g=2$ and $L=22$. The data shows good agreement with the Wigner-Dyson distribution for the GOE. (c) Half-chain von Neumann entanglement entropy of eigenstates for $g=2$ in the most symmetric sector with $L=18$. While at any energy $E$ most eigenstates display a similar value of $S_{L/2}$, there are still a fair number of outliers visible. }
\label{sup_fig:fig_spectrum_SM}%
\end{figure*}

In Fig.~\ref{sup_fig:fig_spectrum_SM}, we study the spectral statistics of the (perturbed) East-West model in Eq.~\ref{eq:px_model_SM} to show that it is chaotic and that all symmetries have been properly identified and resolved. To do so, we focus on the statistics of the energy level spacings $s_j=E_{j+1}-E_j$ and of their ratios $r_j=\frac{\min(s_{j-1},s_j)}{\max(s_{j-1},s_j)}$. For a fully chaotic model, the mean $\langle r \rangle$ of the level-spacing ratios shown on panel (a) should converge towards $\langle r \rangle\approx 0.53$ expected for a random matrix belonging to the Gaussian Orthogonal Ensemble (GOE). Meanwhile, integrability would be indicated by a value $\approx 0.39$, and any unresolved symmetry would lead to a value in between these two cases. Once all symmetries described above have been resolved, we find good agreement with the GOE predictions for both $g=2$ and $g\neq2$. Focusing on the most symmetric sector ($p=0$, $k=0$, $i=1$) for $g=2$, we show additional signs of chaos in panels (b) and (c) through the full level-spacing statistics and the entanglement entropy of eigenstates. The former matches the Wigner-Dyson distribution while the latter shows the individual eigenstates clustering around a smooth curve $S_{L/2}(E)$. Nonetheless, a few eigenstates show some anomalously low entanglement entropy for their energy, as is typical in models displaying weak ergodicity breaking.

\section{Late time relaxation}\label{SM:late_time_relaxation}
\begin{figure}[htbp]
\centering
\includegraphics[width=4.5in]{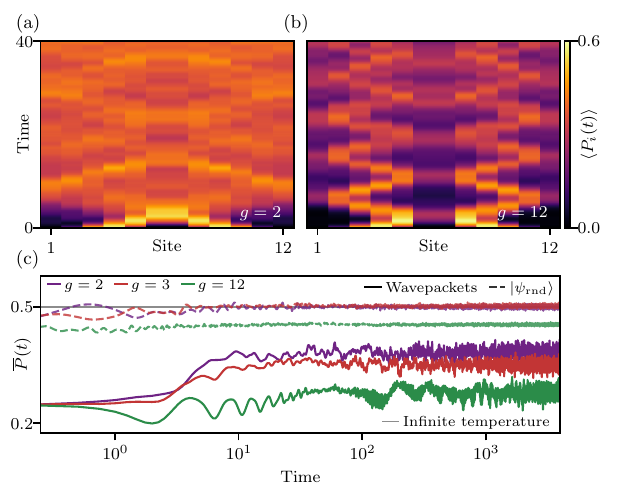}
\caption{Relaxation at finite quasiparticle density under $H_\mathrm{PX}^{(g)}$ on a periodic chain with $L=12$ sites. The initial state $\ket{\chi_{\boldsymbol\alpha,\boldsymbol\beta}}$ contains two compact packets with $\boldsymbol\alpha=(6,3,-\pi/2)$ and $\boldsymbol\beta=(18,3,\pi/2)$, obtained by retaining the first twelve sites of the three-packet state in Fig.~5 of the main text. (a),(b) Local down-projector density $\langle P_i(t)\rangle$ for $g=2$ and $g=12$ up to $t=40$. (c) Mean down-projector density $\overline P(t)=L^{-1}\sum_i\langle P_i(t)\rangle$ for the two-packet state (solid lines) and a random product state (dashed lines), for $g=2,3,12$ up to $t=4000$. The horizontal gray line marks the infinite-temperature value $1/2$, used as an approximate reference. The random state consists of the first twelve local spinors specified in Section~\ref{SM:rnd_product_state}.}
\label{sup_fig:late_time_relaxation}
\end{figure}

To assess the system-size dependence of the late-time dynamics shown in Fig.~5 of the main text, we repeat the simulations on a smaller periodic chain with $L=12$ sites. We retain the first twelve sites of each initial state. Both states factorize across this cut, so the smaller system starts from the corresponding pure state. The wavepacket state contains two separated quasiparticles with the same compact envelopes as before, preserving the quasiparticle density, $2/12=3/18$. The random product state uses the first twelve local spins of the realization specified in Section~\ref{SM:rnd_product_state}. Figure~\ref{sup_fig:late_time_relaxation} shows their evolution for the same couplings and time intervals as in the main text.

For the random product state at $g=2$ and $g=3$, the late-time mean densities are close for the two sizes and remain near the infinite-temperature reference $1/2$. This suggests that the late-time mean density is already only weakly dependent on system size for these parameters. At $g=12$, however, the difference between the two sizes remains appreciable. The states with embedded quasiparticles show a pronounced size dependence for all three couplings, and their late-time fluctuations are larger on the smaller chain. Thus, this comparison does not establish whether the $L=18$ results have converged with system size. Retaining only the first twelve sites also changes the relative numbers of left- and right-moving quasiparticles and the energy density of the random product state, which should be kept in mind when comparing the two systems. Evolving larger systems over such long times is computationally demanding because of the growth of the Hilbert space. A systematic study of the thermodynamic limit and of the persistence of these nonthermal signatures is therefore left for future work.

\section{Parent Hamiltonians}\label{SM:parent_hamiltonians}
The chiral dynamics of the $H_{\mathrm{PX}}$ model can also be related to the local parent Hamiltonian framework of Ref.~\cite{gioia_distinct_2025}, developed for the $\ket{W}=L^{-1/2}\sum_j\ket{0\cdots0\,1_j\,0\cdots0}$ state. To make this connection explicit, we split the Hamiltonian as $H_{\mathrm{PX}}=H_{\mathrm{hop}}+H_1$, with
\begin{equation}
H_{\mathrm{hop}} = \sum_i
\big(
n_{i-1} P_i X_{i+1}
+
X_i P_{i+1} n_{i+2}
\big),
\qquad
H_1 = \sum_i
\big(
P_{i-1}P_i X_{i+1}
+
X_{i-1}P_iP_{i+1}
\big),
\label{eq:ham_split}
\end{equation}
with $n_i=\ket{1}\bra{1}$. Within the tight-binding manifold $\mathcal{H}_{\mathrm{TB}}$, the term $H_{\mathrm{hop}}$ acts exactly as the single-particle tight-binding Hamiltonian on $2L$ sites, defined in Eq.~(5) of the main text, without generating any leakage states $\ket{L_m}$. Thus $\mathcal{H}_{\mathrm{TB}}$ is an exact invariant subspace of $H_{\mathrm{hop}}$, while $H_1$ couples it to the rest of the Hilbert space and turns the dynamics into a genuine many-body problem. This splitting of the Hamiltonian makes a parent Hamiltonian structure of $H_1$ with respect to our exact QMBSs $\{\ket{vac},\ket{\pi/2},\ket{-\pi/2}\}$ apparent because the grouped local terms $h_{1,i}=P_{i-1}P_iX_{i+1}+X_{i-1}P_iP_{i+1}$ annihilate these states. For $\ket{\pm\pi/2}$ this follows from the destructive interference between the two processes creating the same three-zero configuration $\ket{L_i}$. Following the procedure of Ref.~\cite{gioia_distinct_2025}, one can then form chirally propagating droplets by embedding a finite region of a locally annihilated state into the vacuum. In our case, we can build the states $\ket{\psi_\pm(A)}=\ket{\pm\pi/2}_{[A]}\otimes\ket{vac}_{[A^c]}$, where $A$ is a connected region of the chain and $\ket{\pm\pi/2}_{[A]}$ denotes the corresponding momentum state restricted to this interval. Under $H_{\mathrm{hop}}$, this state evolves as a wavepacket centered around $k=\pm\pi/2$ with an approximately rectangular real-space envelope under perfect single-particle tight-binding dynamics. This leads to chiral propagation of the droplet with the direction set by the sign of the momentum. The term $H_1$ spoils the perfect tight-binding dynamics by coupling the state to $\mathcal{H}_\perp$. However, since it locally annihilates the states $\{\ket{vac},\ket{\pi/2},\ket{-\pi/2}\}$, it only acts nontrivial between the boundaries of $A$ and $A^c$. Therefore, sufficiently large droplets are expected to show chiral propagation, while the boundaries start to melt under dynamics of $H_{\mathrm{PX}}$. Replacing the rectangular envelope of the droplet by an approximately Gaussian form, the violation of the ideal tight-binding dynamics spreads more evenly across the wavepacket, reduces sharp boundary effects and suppresses sidelobes in momentum space. From this parent Hamiltonian point of view, one expects chiral propagation of large defects, while the coherent propagation of the rather small defects shown in the main text is more surprising. The connection raises questions with regard to the parent Hamiltonian framework. Does a larger set of locally annihilated eigenstates generally allow for richer nonthermal dynamics? Can different choices of eigenstates lead to qualitatively different propagating droplets or other dynamical signatures?

\section{Rydberg implementation}\label{SM:rydberg}
We use the same spin convention as in the main text: $\ket{0}$ is the atomic ground state, $\ket{1}=\ket{r}$ is the Rydberg state, $P_j=\ket{0}_j\bra{0}$, and $n_j=\ket{1}_j\bra{1}=1-P_j$. In this convention the desired effective Hamiltonian is $J[H_{\mathrm{PX}}+(g-2)H_{\mathrm{PXP}}]$, where $J$ sets the time scale. We set $\hbar=1$. The construction uses the separation of local transition frequencies by the Rydberg interaction, as in blockade and facilitation schemes discussed in the main text.

For equally spaced atoms, let $V$ be the nearest-neighbor interaction. In a frame rotating at the bare atomic transition, the driven Hamiltonian is
\begin{equation}
H(t)=V\sum_j n_jn_{j+1}+H_{\mathrm{tail}}
+\frac12\sum_j\left[\Omega(t)\sigma_j^+
+\Omega^*(t)\sigma_j^-\right],
\label{eq:rydberg_driven}
\end{equation}
where $\sigma_j^+=\ket{1}_j\bra{0}$ and $H_{\mathrm{tail}}=\sum_{i<j,\,|i-j|>1}V_{ij}n_in_j$ collects the interactions beyond nearest neighbors. Taking the interaction picture with respect to $V\sum_jn_jn_{j+1}$ gives
\begin{align}
\sigma_{j,I}^+(t)
={}&P_{j-1}\sigma_j^+P_{j+1}\nonumber\\
&+e^{iVt}\left(n_{j-1}\sigma_j^+P_{j+1}
+P_{j-1}\sigma_j^+n_{j+1}\right)\nonumber\\
&+e^{2iVt}n_{j-1}\sigma_j^+n_{j+1}.
\label{eq:rydberg_dressed}
\end{align}
Thus the transition is shifted by $0$, $V$, or $2V$ when zero, one, or two neighboring atoms are excited. A Rabi drive at the bare atomic transition frequency is therefore resonant only when both neighbors are unexcited, giving the usual PXP blockade term. We add a second Rabi drive at the bare transition frequency shifted by $V$, which is resonant only when exactly one neighbor is excited. The two drives take the form
\begin{equation}
\Omega(t)=\Omega_0+\Omega_1e^{-iVt},
\label{eq:rydberg_drive}
\end{equation}
which allows the zero- and one-neighbor processes to be controlled independently; the two-neighbor transition remains off resonance. Taking real amplitudes with the same phase, the rotating-wave approximation gives
\begin{align}
H_{\mathrm{eff}}
={}&\frac{\Omega_0}{2}\sum_jP_{j-1}X_jP_{j+1}\nonumber\\
&+\frac{\Omega_1}{2}\sum_j\left(n_{j-1}X_jP_{j+1}
+P_{j-1}X_jn_{j+1}\right)+H_{\mathrm{tail}}\nonumber\\
={}&\frac{\Omega_1}{2}H_{\mathrm{PX}}
+\left(\frac{\Omega_0}{2}-\Omega_1\right)H_{\mathrm{PXP}}
+H_{\mathrm{tail}}.
\label{eq:rydberg_effective}
\end{align}
In the last line we used $n_j=1-P_j$ and relabeled the sum in $H_{\mathrm{PX}}$. Consequently,
\begin{equation}
\Omega_1=2J,\qquad \Omega_0=2Jg
\label{eq:rydberg_amplitudes}
\end{equation}
yield the target Hamiltonian up to the residual interactions. The bare-transition amplitude must account for the factor of two already contributed by $H_{\mathrm{PX}}$ when both neighbors are in $\ket{0}$. Equivalently, interchanging the two spin labels maps this construction to tones at $V$ and $2V$, selecting one or two excited neighbors, with amplitudes $2J$ and $2Jg$.

The approximation requires $|V|\gg |\Omega_0|,|\Omega_1|$. The leading off-resonant corrections have scale $\Omega_{\max}^2/|V|$, while $H_{\mathrm{tail}}$ is not removed by this averaging. Large $g$ therefore makes the required separation of drive and interaction scales more demanding.

\section{Data for random product state}
\label{SM:rnd_product_state}
The random initial product state shown in Fig.~\ref{fig:scarred_dynamics} is defined on $L=24$ sites and was obtained by independently drawing each local spin state uniformly from the Bloch sphere. For the comparison on $L=18$ sites in Fig.~5 of the main text, we retain the first $18$ local spinors of the same realization:
\begin{equation}
\ket{\psi_\mathrm{rnd}}
=
\bigotimes_{i=1}^{L}
\left[
\cos\!\left(\frac{\theta_i}{2}\right)\ket{0}_i
+
e^{i\phi_i}
\sin\!\left(\frac{\theta_i}{2}\right)\ket{1}_i
\right].
\label{eq:random_product_state}
\end{equation}
The angles were sampled independently by drawing $\cos(\theta_i)$ uniformly from $[-1,1]$ and $\phi_i$ uniformly from $[0,2\pi)$. The code uses the full-precision realization generated with seed $12$, drawing the two angles successively at each site. The values below are rounded for display:
\begin{align}
(\cos\theta_1,\ldots,\cos\theta_{24})
={}&
(
-0.49835108,
-0.62135923,
-0.30022152,
0.34089149,
0.79261875,
-0.99434594,
\nonumber\\
&\quad
-0.78629745,
-0.16620792,
-0.06370682,
-0.48245782,
0.34102095,
0.84562175,
\nonumber\\
&\quad
-0.87128861,
0.29848074,
-0.18380306,
0.58594015,
0.55768018,
-0.73129653,
\nonumber\\
&\quad
-0.95954255,
-0.72982427,
-0.16198590,
-0.65950746,
0.82904917,
0.20056958
),
\label{eq:random_product_cos_theta_values}
\\
(\phi_1,\ldots,\phi_{24})
={}&
(
5.94862418,
1.12652116,
1.44853337,
0.72306508,
5.39179288,
3.40213223,
\nonumber\\
&\quad
1.62077881,
2.85015415,
5.82775931,
1.18054901,
5.94778073,
5.53077387,
\nonumber\\
&\quad
5.88543533,
5.47614688,
1.37846761,
4.15717199,
1.26508605,
4.79799794,
\nonumber\\
&\quad
5.94138431,
3.77060413,
2.03509263,
4.90328975,
4.57963396,
4.47019650
).
\label{eq:random_product_phi_values}
\end{align}

\endgroup
\end{document}